\documentclass[12pt]{article}
\usepackage{amsmath,amssymb,amsfonts,amsthm}
\usepackage[mathscr]{eucal}
\usepackage[titletoc,title]{appendix}
\usepackage{amssymb}
\usepackage{amsthm}
\theoremstyle{plain}

\newcommand{\be}{\begin{equation}}
\newcommand{\ee}{\end{equation}}
\newcommand{\bea}{\begin{eqnarray}}
\newcommand{\eea}{\end{eqnarray}}
\newcommand{\nn}{\nonumber \\}
\newcommand{\lb}{\label}
\newcommand{\p}[1]{(\ref{#1})}
\numberwithin{equation}{section}

\begin{document}

\begin{titlepage}

\vspace*{0.2cm}

\renewcommand{\thefootnote}{\star}
\begin{center}

{\LARGE\bf Gauged spinning models   }\\

\vspace{0.5cm}

{\LARGE\bf with deformed supersymmetry}\\

\vspace{1.5cm}
\renewcommand{\thefootnote}{$\star$}

{\large\bf Sergey~Fedoruk} \footnote{{\it On leave of absence from V.N.\,Karazin Kharkov National University, Ukraine}},
\quad {\large\bf
Evgeny~Ivanov
}
 \vspace{0.5cm}

{ \it Bogoliubov Laboratory of Theoretical Physics, JINR,}\\
{\it 141980 Dubna, Moscow region, Russia} \\
\vspace{0.1cm}

{\tt fedoruk@theor.jinr.ru, \; eivanov@theor.jinr.ru
}\\
\vspace{0.7cm}

\end{center}
\vspace{0.2cm} \vskip 0.6truecm \nopagebreak

   \begin{abstract}
\noindent
New models of the  ${\rm SU}(2|1)$ supersymmetric mechanics based on gauging the systems with dynamical ${\bf (1,4,3)}$
and  semi-dynamical ${\bf (4,4,0)}$ supermultiplets  are presented. We propose a new version of ${\rm SU}(2|1)$ harmonic superspace  approach which makes it possible to construct the Wess-Zumino term for interacting ${\bf (4,4,0)}$ multiplets.
A new ${\cal N}=4$ extension of $d=1$ Calogero-Moser multiparticle system is obtained by gauging the ${\rm U}(n)$
isometry of matrix ${\rm SU}(2|1)$ harmonic superfield model.

\end{abstract}

\vspace{1cm}

\phantom{Keywords: }

\newpage

\end{titlepage}

\setcounter{footnote}{0}

\setcounter{equation}0
\section{Introduction}

In recent papers \cite{IS14a,IS14b,IS15} there was initiated the systematic study of the models of deformed ${\cal N}=4$ supersymmetric mechanics
with ${\rm SU}(2|1)$ as a substitute of the standard ``flat'' ${\cal N}=4, d=1$ superalgebra.
Earlier examples of ${\rm SU}(2|1)$ supersymmetric $d=1$ models have been pioneered in  \cite{Sm,BellNer}.
The higher-dimensional systems with curved rigid supersymmetry based on the supergroup  ${\rm SU}(2|1)$ and
its central extension were studied in \cite{FesSei,DumFesSei,SamSor}.

The centrally-extended superalgebra $\hat{su}(2|1)$ \cite{IS14a,IS14b,IS15} is spanned  by the fermionic generators $Q^{i}$
and $\bar{Q}_{i}=(Q^{i})^\dagger\,, \;i=1,2\,$, satisfying
\begin{equation}\label{susy}
\lbrace Q^{i}, \bar{Q}_{k}\rbrace = 2m I^i_k + 2\delta^i_k\left(H-mF\right) ,\qquad
\lbrace Q^{i}, Q^{k}\rbrace =\lbrace \bar{Q}_{i}, \bar{Q}_{k}\rbrace =0\,.
\end{equation}
The generator $H=H^\dagger$ commutes with all other generators and can be interpreted as an operator central charge.
The ${\rm SU}(2)_{\rm int}$ generators $I^i_k=(I^k_i)^\dagger$ and the ${\rm U}(1)_{\rm int}$ generator $F=F^\dagger$,
\begin{equation}\label{u2}
\left[I^i_j,  I^k_l\right] = \delta^k_j I^i_l - \delta^i_l I^k_j\,,\qquad
\left[I^i_j,  F\right] =0\,,
\end{equation}
possess the non-vanishing commutators with supercharges
\begin{equation}\label{u2-q}
\left[I^i_j, Q^{k}\right] = \delta^k_j Q^{i} - \frac{1}{2}\,\delta^i_j Q^{k}\, ,\qquad
\left[I^i_j, \bar{Q}_{l}\right] = -\delta^i_l\bar{Q}_{j} +\frac{1}{2}\,\delta^i_j\bar{Q}_{l}\,,
\end{equation}
\begin{equation}\label{u2-i}
\left[F, Q^{k}\right]=\frac{1}{2}\,Q^{k}\,,\qquad \left[F, \bar{Q}_{l}\right]=-\frac{1}{2}\,\bar{Q}_{l}\,.
\end{equation}
Furthermore, the $su(2|1)$ superalgebra has the automorphism group ${\rm SU}(2)_{\rm ext}$ with
the generators $T^i_j=(T^k_i)^\dagger$ which rotate the supercharges
in the precisely same manner as the internal ${\rm SU}(2)_{\rm int}$ generators $I^i_j$ do:
\begin{equation}\label{u2-t-q}
\left[T^i_j, Q^{k}\right] = \delta^k_j Q^{i} - \frac{1}{2}\,\delta^i_j Q^{k}\, ,\qquad
\left[T^i_j, \bar{Q}_{l}\right] = -\delta^i_l\bar{Q}_{j} +\frac{1}{2}\,\delta^i_j\bar{Q}_{l}\,.
\end{equation}
The  ${\rm SU}(2)_{\rm ext}$ generators rotate, in the same way, the indices of
the ${\rm SU}(2)_{\rm int}$ generators $I^i_j$, so these two ${\rm SU}(2)$ groups form a semi-direct product
\begin{equation}\label{u2-u2}
\left[T^i_j,  I^k_l\right] = \delta^k_j I^i_l - \delta^i_l I^k_j\,.
\end{equation}

In \cite{IS14a,IS14b} the  ${\rm SU}(2|1)$ invariant one-particle $d=1$ models were constructed, proceeding from the superfield formalism
based on the superspace with the coordinates $ \left(t,\theta_{k}, \bar{\theta}^{k}\right)$ $\bar\theta^i=(\overline{\theta_i^{\!\!\phantom{k}}})$.
These coordinates are related to the ${\rm SU}(2|1)$ coset representative $\exp\left\{it H+\vartheta_k Q^{k}+\bar\vartheta^k \bar Q_{k} \right\}$
via the substitutions $\vartheta_i=\left( 1+\frac23\,m\, \theta_k\bar\theta^k\right)\theta_i\,$.
$\bar\vartheta^i=\left( 1+\frac23\,m\, \theta_k\bar\theta^k\right)\bar\theta^i$.
The fermionic ${\rm SU}(2|1)$  transformations are realized on them as
\be\label{CBtr}
\delta t=i\left(\epsilon_k\,\bar{\theta}^k+\bar{\epsilon}^k\,\theta_k  \right)\,,\qquad
\delta\theta_{i}=\epsilon_{i}+2m\,\bar{\epsilon}^k\,\theta_k \,\theta_{i}\,,\qquad
\delta\bar{\theta}^{i} =\bar{\epsilon}^{i}-2m\,\epsilon_k\,\bar{\theta}^k\,\bar{\theta}^{i}\,.
\ee

As a further step, in \cite{IS15} there was considered the ``minimal'' complex harmonic coset
\be \label{HB}
\frac{\{H, Q^\pm, \bar Q^\pm, F, I^{\pm\pm}, I^0,  T^{\pm\pm}, T^0\}}{\{F, I^{++}, I^0, I^{--} - T^{--}, T^0\}} \sim
\left(t_{A}, \theta^\pm, \bar\theta^\pm, w^\pm_i\right)\equiv \zeta_H\, ,
\ee
where
\be\label{I-pm}
I^{++}\equiv I^1_2\,,\qquad I^{--}\equiv I^2_1\,,\qquad I^{0}\equiv I^1_1-I_2^2 =2I^1_1\,,
\ee
\be\label{T-pm}
T^{++}\equiv T^1_2\,,\qquad T^{--}\equiv T^2_1\,,\qquad T^{0}\equiv T^1_1-T_2^2 =2T^1_1\,.
\ee
\be\label{Q-pm}
Q^+ \equiv Q^1 \,,\qquad Q^- \equiv Q^2 \,,\qquad \bar{Q}^-\equiv \bar{Q}_1 \,,\qquad \bar{Q}^+\equiv -\bar{Q}_2 \,.
\ee
This ${\rm SU}(2|1)$ harmonic approach, as a deformation of the analogous formalism in ${\cal N}=4$ supersymmetric mechanics \cite{IvLech},
have provided additional opportunities to build new ${\rm SU}(2|1)$  models, in particular those associated
with the multiplet $({\bf 4, 4, 0})$ and its ``mirror'' counterpart.
As was pointed out in \cite{IS14a,IS14b,IS15} (see also  \cite{IS16}), many issues of ${\cal N}=4$ supersymmetric mechanics still
await their ${\rm SU}(2|1)$ generalization. The list includes the ${\cal N}=4$ supersymmetric Calogero-like systems, the gauging
procedure in superspace, coupling to the background gauge fields, etc.
In the framework of ${\cal N}=4$ supersymmetric mechanics, all these topics were found to be tightly interrelated.
E.g., the Wess-Zumino (WZ) type actions describe the interaction of the proper $d=1$ supermultiplets with external gauge fields \cite{IvLech}.
The actions of the same type describe semi-dynamical degrees of freedom \cite{FIL09,FIL10},
the use of which proved to be of pivotal importance for constructing the many-particle supersymmetric $d=1$ systems \cite{FIL08}
(see also the review \cite{FIL12}).
Additional important technical ingredients of the ${\cal N}=4$ model-building which essentially exploit the WZ type $d=1$ actions
are the pure gauge ``topological'' multiplet and the superfield gauging procedure relating diverse models \cite{DI06,DI07}.

In this paper we construct new models of the ${\cal N}=4$ deformed supersymmetric mechanics that make use of a few different types
of ${\rm SU}(2|1)$ supermultiplets: dynamical, semi-dynamical and pure gauge supermultiplets.
The outcome are new ${\rm SU}(2|1)$-invariant one-particle model with spinning degrees of freedom, as well as new ${\rm SU}(2|1)$ superextension
of the Calogero-Moser multi-particle system.

The harmonic superspace  \p{HB}  is not directly applicable for tackling these tasks.
The main problem roots in the algebra of the covariant constraints  to be imposed on the relevant
harmonic superfields $\Psi$ for singling out various irreducible ${\rm SU}(2|1)$ multiplets.
The Grassmann analyticity conditions in the harmonic superspace  \p{HB}
(specifically, ${\cal D}^+\Psi=0$, $\bar{\cal D}^+\Psi=0$) necessarily entail the harmonic condition (specifically, ${\cal D}^{++}\Psi=0$).
However, such harmonic constraints turn out to be too strong if we wish to describe some supermultiplets in the harmonic approach, e.g.
the ``topological'' gauge multiplet  which is the main object of the $d=1$ gauging \cite{DI06,DI07} efficiently exploited in
refs. \cite{FIL08,FIL09,FIL10,FIL12}.
As we will see, the only way around is to pass to an extended ${\rm SU}(2|1)$ harmonic superspace involving two sets of harmonic variables:
those associated with the group ${\rm SU}(2)_{\rm int}$ and those parametrizing the external automorphism group
${\rm SU}(2)_{\rm ext}$.

In Sect.\,2 we introduce new harmonic superspace with two sets of harmonic variables, including the standard (unitary) harmonics on ${\rm SU}(2)_{\rm ext}$.
As a result, we gain an opportunity to perform a gauging procedure and define interacting dynamical and semi-dynamical multiplets.
In Sect.\,3 we construct the system of dynamical ${\bf (1,4,3)}$ multiplet interacting with a semi-dynamical ${\bf (4,4,0)}$ multiplet.
This coupling is used to define the WZ term for the ${\bf (4,4,0)}$ multiplet, which, as was noticed in \cite{IS15}, is impossible in the framework
of the harmonic superspace \p{HB}. The gauging procedure relevant to this ${\rm SU}(2|1)$ invariant system is described
in  Sect.\,4. In Sect.\,5 we present a matrix generalization of  the ${\rm SU}(2|1)$ invariant model
with dynamical, semi-dynamical and pure gauge supermultiplets. When reduced on shell, it describes
${\rm SU}(2|1)$ supersymmetrization of the Calogero-Moser multi-particle system \cite{C,Moser}, with the mass
specified by the deformation parameter $m$ of the $su(2|1)$ algebra. Sect.\,6 contains the concluding remarks. In Appendix we present
the ``master'' ${\rm SU}(2|1)$ harmonic formalism which yields the settings developed in \cite{IS15} and in Sect.\,2 of the present paper
upon two different reductions with respect to the extra harmonic variables.

\setcounter{equation}0
\section{${\rm SU}(2|1)$ harmonic superspace revisited}

As opposed to the ``minimal'' harmonic coset
\p{HB}, we will use now the coset
\be\label{HB-new}
\hat{\zeta}_H=\left(t_{A}, \theta^\pm, \bar\theta^\pm,  u^\pm_i, z^{++} \right) \quad\sim\quad \frac{\{H, Q^\pm,
\bar Q^\pm, F, I^{\pm\pm}, I^0,  T^{\pm\pm}, T^0\}}{\{F, I^{++}, I^0, T^0\}}
\,,
\ee
where the variables
\be\label{u-harm}
u^\pm_i\,, \qquad u^{+i}u^-_i =1\,,
\qquad u^+_iu^-_k - u^+_k u^-_i = \varepsilon_{ik}
\ee
are the standard unitary harmonics
on the coset ${\rm SU}(2)_{\rm ext}/{\rm U}(1)_{\rm ext}\sim S^2$ \cite{HSS0}, while the coordinate $z^{++}$ is associated with the generator $I^{--}$.
The elements of this coset are defined as
\be\label{coset-repr}
g_H=e^{i\left(\xi T^{++}+\bar\xi T^{--} \right) } \exp\left\{z^{++}I^{--} \right\}
\exp\left\{it_A H-\theta^+ Q^{-}+\bar\theta^+ \bar Q^{-} \right\}\exp\left\{\theta^- Q^{+}-\bar\theta^- \bar Q^{+} \right\} \,,
\ee
where $e^{i(\, \xi \tau^{++}+\bar\xi \tau^{--})  } =(u^\pm_i)$,
$\tau^{\pm\pm}=\frac{1}{2}(\tau^1\pm i\tau^2)$, $\tau^p$, $p=1,2,3$ are the Pauli matrices, and we use the notations \p{I-pm}, \p{T-pm}, \p{Q-pm}.

The relation with the standard ${\rm SU}(2|1)$  superspace coordinates is given by
\be\label{var-new}
\begin{array}{c}
t_{A} = t + i\left(\theta^{+}\bar{\theta}^{-} + \theta^{-}\bar{\theta}^{+}\right)\,,\\[5pt]
\theta^{-}=\theta^{i}w^{-}_{i}  ,\qquad \theta^{+}=
\theta^{i} w^{+}_{i} \left(1 + m\,\theta^{k} w^{-}_{k}\bar{\theta}^{l} w^{+}_l \right),\\[5pt]
\bar{\theta}^{-}= \bar{\theta}^{k} w^{-}_k ,\qquad
\bar{\theta}^{+}=\bar{\theta}^{k} w^{+}_k
\left(1 - m\,\theta^{k} w^{+}_{k}\bar{\theta}^{l} w^{-}_l\right),
\end{array}
\ee
where $w^\pm_i$ are the non-unitary harmonics which define the ``minimal'' complex harmonic coset
\p{HB} and are related to the harmonics \p{u-harm} as \cite{HSS,GIO}
\be \label{w-v}
w^+_i=u^+_i +z^{++}u^-_i\,,\qquad w^-_i=u^-_i\,, \qquad w^+_iw^-_k - w^+_k w^-_i = \varepsilon_{ik}\,.
\ee
The relations \p{var-new} imply \cite{IS15}
\be\label{var-new1}
\begin{array}{c}
t = t_{A} - i\left(\theta^{+}\bar{\theta}^{-} + \theta^{-}\bar{\theta}^{+}\right)\,,\\[5pt]
\theta^{i}w^{-}_{i}=\theta^{-}  ,\quad \theta^{i} w^{+}_{i}=
\theta^{+} \left(1 - m\,\theta^{-}\bar{\theta}^{+} \right),\qquad
\bar{\theta}^{k} w^{-}_k = \bar{\theta}^{-},\quad \bar{\theta}^{k} w^{+}_k=\bar{\theta}^{+}
\left(1 + m\,\theta^{+}\bar{\theta}^{-}\right).
\end{array}
\ee

The fermionic ${\rm SU}(2|1)$ transformations induced by the left shifts of the coset representative \p{coset-repr} are written as
\be\label{HB-tr}
\begin{array}{c}
\delta t_{A} = 2 i \left( \epsilon^{-}\bar{\theta}^{+}-\bar{\epsilon}^{-}\theta^{+} \right)\,,\\[5pt]
\delta \theta^{+}= \epsilon^{+}+\epsilon^{-}\left(z^{++}- m\,\theta^{+}\bar{\theta}^{+}\right), \qquad
\delta \bar{\theta}^{+}= \bar{\epsilon}^{+} +\bar\epsilon^{-}\left(z^{++}+ m\,\theta^{+}\bar{\theta}^{+}\right),\\[5pt]
\delta \theta^{-}= \epsilon^{-}+ 2m\,\bar{\epsilon}^{-}\theta^{-} \theta^{+}, \qquad
\delta \bar{\theta}^{-}= \bar{\epsilon}^{-} + 2m\,\epsilon^{-}\bar{\theta}^{-}\bar{\theta}^{+},\\[5pt]
\delta z^{++}= m\left(\epsilon^{+}\bar{\theta}^{+} + \bar{\epsilon}^{+} \theta^{+} \right)
+m\,z^{++}\left(\epsilon^{-}\bar{\theta}^{+} + \bar{\epsilon}^{-} \theta^{+} \right)\,,\\[5pt]
\delta u^\pm_{i}=0\,,
\end{array}
\ee
where
\bea
\epsilon^{\pm} = \epsilon^{i}u^{\pm}_{i}\,, \qquad \bar{\epsilon}^{\pm} = \bar{\epsilon}^{k} u^{\pm}_k\,.
\eea
It follows from the transformations \p{HB-tr} that the ${\rm SU}(2|1)$ harmonic superspace contains the analytic harmonic
subspace parametrized by the reduced coordinate set
\be \lb{AnalSubs}
\hat{\zeta}_{A} =\left( t_{A}, \bar{\theta}^{+}, \theta^{+},  u^{\pm}_i, z^{++}\right),
\ee
which is closed under the action of ${\rm SU}(2|1)$. It can be identified with the supercoset
\bea
\hat{\zeta}_{A} \quad\sim\quad \frac{\{H, Q^\pm, \bar Q^\pm, F, I^{\pm\pm}, I^0,  T^{\pm\pm}, T^0\}}{\{Q^+, \bar Q^+, F, I^{++}, I^0, T^0\}}    \,.
\lb{AnalCos}
\eea
The transformations \p{HB-tr} rewritten through harmonics $w^\pm_i$ defined in \p{w-v}
take just the form given in \cite{IS15}
\be\label{HBtr}
\begin{array}{c}
\delta t_{A} = 2 i \left( \eta^{-}\bar{\theta}^{+} - \bar{\eta}^{-}\theta^{+} \right),\\[5pt]
\delta \theta^{+}= \eta^{+}- m\,\eta^{-}\theta^{+}\bar{\theta}^{+}\,, \qquad
\delta \bar{\theta}^{+}= \bar{\eta}^{+} + m\,\bar{\eta}^{-}\theta^{+}\bar{\theta}^{+}\,,\\[5pt]
\delta \theta^{-}= \eta^{-}+ 2m\,\bar{\eta}^{-}\theta^{-} \theta^{+}, \qquad
\delta \bar{\theta}^{-}= \bar{\eta}^{-} + 2m\,\eta^{-}\bar{\theta}^{-}\bar{\theta}^{+},\\[5pt]
\delta w^+_{i}=m\left(\eta^{+}\bar{\theta}^{+} + \bar{\eta}^{+}\theta^{+}  \right)w^-_{i},\qquad \delta w^-_{i}=0\,,
\end{array}
\ee
where $\eta^{\pm} = \epsilon^{i}w^{\pm}_{i}$, $\bar{\eta}^{\pm} = \bar{\epsilon}^{i} w^{\pm}_i$. The extra coordinate $z^{++}$
transforms in this basis as
\be
\delta z^{++} = m\left(\eta^{+}\bar{\theta}^{+} + \bar{\eta}^{+}\theta^{+}  \right). \lb{Trnsfz++}
\ee

Applying the routine coset techniques to the coset \p{HB-new} (see, for example, \cite{IS14a})
we derive the following expressions for the covariant derivatives
\be\lb{Dt}
{\cal D}_{t_A} = \partial_{\,t_A}=\frac{\partial}{\partial t_{A}}\,,
\ee
\be\lb{D-}
\begin{array}{rcl}
{\cal D}^{-} &=& {\displaystyle -\frac{\partial}{\partial \theta^+} - 2i \,\bar\theta^-\partial_{t_A}
-m\,\bar{\theta}^{-} \theta^-\frac{\partial}{\partial \theta^-}
+ m\,\bar\theta^+\frac{\partial}{\partial z^{++}}  \, + m\,\bar\theta^- \left(\tilde{I}^0 +2\tilde{F}\right)\,, }\\ [7pt]
\bar{{\cal D}}^{-} &=& {\displaystyle \frac{\partial}{\partial \bar\theta^+} - 2i\, \theta^- \partial_{t_A}
+m\,\theta^{-}\bar\theta^-\frac{\partial}{\partial \bar\theta^-}-
m\,\theta^+\frac{\partial}{\partial z^{++}}  \, - m\, \theta^- \left(\tilde{I}^0 -2\tilde{F}\right)\,, }
\end{array}
\ee

\be\lb{D+}
\begin{array}{rcl}
{\cal D}^{+} &=& {\displaystyle  \frac{\partial}{\partial \theta^-}
- m\,\bar\theta^- \tilde{I}^{++} \,,}\\ [7pt]
\bar{{\cal D}}^{+} &=&  {\displaystyle -\frac{\partial}{\partial \bar\theta^-}
+ m\, \theta^- \tilde{I}^{++}\,,}
\end{array}
\ee

\be\lb{D++}
{\cal D}_z^{--} = {\displaystyle \frac{\partial}{\partial z^{++}}
+ 2i \,\theta^{-}\bar\theta^{-}\partial_{t_A} + m\left(\theta^{+} \bar\theta^{-}-\theta^{-} \bar\theta^{+}\right)\frac{\partial}{\partial z^{++}}
+ \theta^{-}\frac{\partial}{\partial \theta^+} +
\bar\theta^{-}\frac{\partial}{\partial \bar\theta^{+}} }- 2 m\,\theta^{-} \bar\theta^{-}\tilde{F}
\,,
\ee

\be\lb{cov-D--}
{\cal D}^{--} =\,\partial^{--}_{u}
+ 2i \,\theta^{-}\bar\theta^{-}\partial_{t_A} + m\left(\theta^{+} \bar\theta^{-}-\theta^{-} \bar\theta^{+}\right)\frac{\partial}{\partial z^{++}}
+ \theta^{-}\frac{\partial}{\partial \theta^+} +
\bar\theta^{-}\frac{\partial}{\partial \bar\theta^{+}}
- 2 m\,\theta^{-} \bar\theta^{-}\tilde{F}\,,
\ee
\be\label{cov-D++}
\begin{array}{rcl}
{\cal D}^{++} &=& {\displaystyle \partial^{++}_{u}
+ 2i \,\theta^+\bar\theta^+\partial_{t_A} + \Big(\theta^{+} + m\,\theta^+\bar\theta^{+} \theta^{-}\Big)\frac{\partial}{\partial \theta^{-}}
+ \Big(\bar\theta^{+}- m\,\theta^+\bar\theta^{+} \bar\theta^{-}\Big)\frac{\partial}{\partial \bar\theta^{-}}
 }
\quad\,\,\,\,\,\,\,\,\,\\ [7pt]
&& {\displaystyle - z^{++}\partial^{0}_{u}- (z^{++})^2\frac{\partial}{\partial z^{++}}
}
\\ [7pt]
&& {\displaystyle + z^{++}\left({\cal D}^0+\tilde{I}^0\right) - 2 m\,\theta^+\bar\theta^{+}\tilde{F}
-m\Big(\theta^{-}\bar\theta^{+} - \theta^{+}\bar\theta^{-} \Big)\tilde{I}^{++}
\,,}
\end{array}
\ee
\be\lb{cov-D0}
{\cal D}^0 \,=\,\partial^0_{u} + 2z^{++}\frac{\partial}{\partial z^{++}}
+ \left(\theta^+\frac{\partial}{\partial \theta^+} + \bar\theta^+\frac{\partial}{\partial \bar\theta^{+}} \right)
- \left(\theta^{-}\frac{\partial}{\partial \theta^{-}} +
\bar\theta^{-}\frac{\partial}{\partial \bar\theta^{-}} \right).\qquad\qquad\quad\quad\,.
\ee

The partial harmonic derivatives in these expressions are defined as
\begin{equation}\label{part-u2}
\partial^{\pm\pm}_{u} = u^\pm_i \frac{\partial}{\partial u^\mp_i} \,,
\quad  \partial^0_{u} = u^+_i \frac{\partial}{\partial u^+_i} - u^-_i \frac{\partial}{\partial u^-_i}\,,
\qquad
[\partial^{++}_{u}, \partial^{--}_{u}] = \partial^0_{u}\,, \quad
[\partial^0_{u}, \partial^{\pm\pm}_{u}] = \pm\,2 \partial^{\pm\pm}_{u}\,,
\end{equation}
and $\tilde{F}$, $\tilde{I}^{0}$, $\tilde{I}^{++}$ are matrix parts of the generators $F$, ${I}^{0}$, ${I}^{++}$
properly acting on the matrix indices of the superfields and the operators. In particular, note the ${\rm U}(1)$ assignments
\begin{equation}\label{I0-op}
\tilde{I}^{0}{\cal D}^{\pm}=\mp{\cal D}^{\pm} \,,
\quad
\tilde{I}^{0}\bar{\cal D}^{\pm}=\mp\bar{\cal D}^{\pm}\,,
\qquad
\tilde{F} {\cal D}^{\pm}=-\frac12\,{\cal D}^{\pm} \,,
\quad
\tilde{F} \bar{\cal D}^{\pm}=\frac12\,\bar{\cal D}^{\pm}\,,
\end{equation}
which will be used below. Note the non-zero commutation relation
\be
[\tilde{I}^0, \tilde{I}^{++}] = 2 \tilde{I}^{++}\,.
\ee
Also, the notable property is
\be
{\cal D}_z^{--} - {\cal D}^{--} =  \frac{\partial}{\partial z^{++}}  -  \partial^{--}_u \,.\lb{D-D}
\ee

The covariant derivatives act on the harmonic superfields $\Psi^{(q)}(t_A, \theta^\pm, \bar\theta^\pm, u^\pm, z^{++}) = \Psi^{(q)}(\hat{\zeta}_H)$
which are assumed to transform under ${\rm SU}(2|1)$ supersymmetry in accord with the general rules of the (super)coset realizations
\be\label{trans-psi}
\delta\Psi^{(q)} =m\left[2\left(\epsilon^-\bar\theta^+ - \bar\epsilon^-\theta^+\right)\tilde F -
\left(\epsilon^-\bar\theta^+ + \bar\epsilon^-\theta^+\right)\tilde I^0 -
\left(\epsilon^-\bar\theta^- + \bar\epsilon^-\theta^-\right)\tilde I^{++} \right]\Psi^{(q)}\,.
\ee
As usual, these superfields are eigenfunctions of the harmonic ${\rm U}(1)$ charge
operator ${\cal D}^0$:
\be\lb{D0-cond}
{\cal D}^{0}\Psi^{(q)} =q\Psi^{(q)}\,.
\ee
We treat the dependence of $\Psi^{(q)}(t_A, \theta^\pm, \bar\theta^\pm, u^\pm, z^{++})$ on two sorts of harmonic variables
in the same way as in \cite{GIO}. Namely, we assume the polynomial dependence on $z^{++}$ and the standard harmonic expansion
in $u^\pm $ \cite{HSS0}.

It is worth pointing out that ${\cal D}^{++}\Psi^{(q)}, \, {\cal D}^+\Psi^{(q)}$ and $\bar{\cal D}^+\Psi^{(q)}$ transform according to the general
superfield rule  \p{trans-psi}, while the ${\rm SU}(2|1)$ variations of ${\cal D}^{--}\Psi^{(q)}$ and ${\cal D}^{-}\Psi^{(q)}, \bar{\cal D}^- \Psi^{(q)}$ exhibit
some deviations from \p{trans-psi}, involving the superfield $\Psi^{(q)}$ itself. However, this subtlety is harmless for our subsequent
consideration.

In what follows  we will mainly limit our study to the harmonic superfields subjected to some additional covariant conditions
\be\lb{I0-cond}
\left({\cal D}^0 + {\tilde I}^0 \right)\Psi^{(q)} = 0 \quad \Rightarrow \quad {\tilde I}^{0}\Psi^{(q)} = -q\Psi^{(q)}\,,
\ee
\be\lb{F-cond}
{\tilde F}\,\Psi^{(q)} = 0\,,
\ee
\be\lb{I++-cond}
{\tilde I}^{++}\Psi^{(q)} = 0\,,
\ee
as well as the constraint
\be\label{D--cond}
\left({\cal D}_z^{--} - {\cal D}^{--}\right) \Psi^{(q)} = 0\,.
\ee

The constraint  \eqref{D--cond} effectively eliminates the dependence of
the harmonic superfields on the variable $z^{++}$
\be\label{psi-sol}
\Psi^{(q)}(t_A, \theta^\pm, \bar\theta^\pm, u^\pm, z^{++}) = e^{\,z^{++}\partial_u^{--}}\Phi^{(q)} (t_A, \theta^\pm, \bar\theta^\pm, u^\pm)\,,
\ee
where $\Phi^{(q)}$ satisfies the condition
\be\lb{D0-cond1}
{D}^{0}\Phi^{(q)} =q\Phi^{(q)}\,,\qquad
{D}^0 = \partial^0_u
+ \theta^+\frac{\partial}{\partial \theta^+} + \bar\theta^+\frac{\partial}{\partial \bar\theta^{+}}
- \theta^-\frac{\partial}{\partial \theta^-} - \bar\theta^-\frac{\partial}{\partial \bar\theta^{-}}
\ee
as a consequence of \p{D0-cond} and has the standard expansion in $u^\pm$.
The superfield solution \p{psi-sol}  can be rewritten as
\be\label{psi-sol1}
\Psi^{(q)}(t_A, \theta^\pm, \bar\theta^\pm, u^\pm, z^{++}) =\Phi^{(q)} (t_A, \theta^\pm, \bar\theta^\pm, w^\pm) = \Phi^{(q)} (\zeta_H)\,,
\ee
where $w^\pm_i$ and $\zeta_H$ were defined in \eqref{w-v} and \eqref{HB}.

The constraint \p{I++-cond} is the self-consistency condition for the covariant definition
of the analytic ${\rm SU}(2|1)$ superfields which live on the analytic subspace \p{AnalSubs}.
This definition amounts to the Grassmann-analyticity constraints
\be\lb{CRsuperf}
{\cal D}^{+}\Psi^{(q)} =\bar{\cal D}^{+}\Psi^{(q)} = 0\,,
\ee
which, due to the relation
\bea
\{{\cal D}^+, \bar{\cal D}^+\} = 2m {\tilde I}^{++} \lb{DbarD}
\eea
following from \p{D+}, necessarily imply \p{I++-cond}.
Similar to \eqref{psi-sol1}, the analytic harmonic superfields are expressed as
\be\label{psi-sol-a}
\Psi^{(q)}(t_A, \theta^+, \bar\theta^+, u^\pm, z^{++}) =e^{\,z^{++}\partial_u^{--}}\Phi^{(q)} (t_A, \theta^+, \bar\theta^+, u^\pm)
=\Phi^{(q)} (t_A, \theta^+, \bar\theta^+, w^\pm)= \Phi^{(q)}(\zeta_A)\,.
\ee
As opposed to the approach of ref. \cite{IS15}, the constraints \p{CRsuperf} and \p{DbarD} by no means require
the condition ${\cal D}^{++}\Psi^{(q)} = 0\,$. Of course the latter can be imposed as an {\it independent} additional constraint, but
it is not necessitated now by the Grassmann analyticity conditions \p{CRsuperf}. The relationship between two alternative
${\rm SU}(2|1)$ harmonic approaches  is  explained in Appendix.

The constraint \p{F-cond} leads to some simplification of
the expressions for other covariant derivatives.
For example, on harmonic superfields obeying the constraints \p{D0-cond} -- \p{D--cond}   the covariant derivative
${\cal D}^{++}$ \p{cov-D++} takes the form
\be\lb{D++-expr}
{\cal D}^{++}\Psi^{(q)} =e^{\,z^{++}\partial_u^{--}}\, D^{++}\,\Phi^{(q)}\,,
\ee
where
\be\lb{D++-short}
{D}^{++} = \partial^{++}_u + 2i \,\theta^+\bar\theta^+\partial_{t_A}
+ \Big(\theta^{+} + m\,\theta^+\bar\theta^{+} \theta^{-}\Big)\frac{\partial}{\partial \theta^{-}}
+ \Big(\bar\theta^{+}- m\,\theta^+\bar\theta^{+} \bar\theta^{-}\Big)\frac{\partial}{\partial \bar\theta^{-}}\,.
\ee
The general transformation law \p{trans-psi} for the superfields subjected to the constraints \p{D0-cond} -- \p{D--cond} is simplified
to  the form
\be\label{trans-psi-s}
\delta\Psi^{(q)} = q m
\left(\epsilon^-\bar\theta^+ + \bar\epsilon^-\theta^+\right) \Psi^{(q)} \,.
\ee

One more comment concerns the possibility to use, along with the harmonic basis $(u^\pm_i, z^{++})$, the basis $(w^\pm_i, z^{++})$
with the non-unitary harmonics. Due to the relation \eqref{w-v}, these two bases are equivalent to each other, while many formulas
and constraints are simplified in the second basis. The dictionary between these bases is as follows
\bea
&& \partial^{++}_u \quad \Rightarrow \quad  \partial^{++}_w + z^{++}\partial_w^0 - (z^{++})^2 \partial_w^{--}\,, \qquad  \partial^{--}_u
\quad \Rightarrow \quad \partial_w^{--}\,, \nn
&&  \partial^{0}_u  \quad \Rightarrow \quad \partial_w^0 - 2z^{++}\partial_w^{--}\,, \qquad \frac{\partial}{\partial z^{++}}
\quad \Rightarrow \quad \frac{\partial}{\partial z^{++}} + \partial_w^{--}\,. \lb{Transit}
\eea
For instance, in the $(w^\pm_i, z^{++})$ basis the constraint \p{D--cond} becomes just the condition of $z^{++}$ independence
\be
\frac{\partial}{\partial z^{++}}\Psi^{(q)} = 0 \quad \Rightarrow \quad \Psi^{(q)} =\Phi^{(q)} (t_A, \theta^\pm, \bar\theta^\pm, w^\pm)\,.
\ee
Its ${\rm SU}(2|1)$ covariance immediately follows from the property $ \delta\frac{\partial}{\partial z^{++}} = 0\,.$ Also, it is instructive to present the
$(w^\pm_i, z^{++})$ form of the pure harmonic part of the covariant derivative ${\cal D}^{++}$ \p{cov-D++}:
\bea
\partial^{++}_u - z^{++}\partial^{0}_u - (z^{++})^2\frac{\partial}{\partial z^{++}} \quad \Rightarrow \quad  \partial^{++}_w
- (z^{++})^2\frac{\partial}{\partial z^{++}}\,.\lb{indPart}
\eea

In construction of the superfield particle actions we will need the expressions for the invariant integration measures over the full harmonic
and the harmonic analytic superspaces \cite{IS15}:
\be\lb{Fullmeas}
d\zeta_H = dw\, dt_{A} \,d\bar{\theta}^{-}d\theta^- d\bar{\theta}^{+}d\theta^{+}
\left(1 +m\, \theta^{+}\bar{\theta}^{-} - m\,\theta^{-}\bar{\theta}^{+} \right)
\ee
and
\be\lb{anmeas}
d\zeta_{A}^{--} = dw\, dt_{A} \,d\bar{\theta}^{+}d\theta^{+}\,,\qquad \delta d\zeta_{A}^{--} =0\,.
\ee

\setcounter{equation}0
\section{Coupling of dynamical multiplet ${\bf (1,4,3)}$ with \\ semi-dynamical multiplet ${\bf (4,4,0)}$}

\subsection{The multiplet ${\bf (1,4,3)}$}

The multiplet ${\bf (1,4,3)}$ is described by the Grassmann-even real superfield $\mathscr{X}$ subjected to the conditions
\eqref{D0-cond}-\eqref{D--cond},
\be\label{x-constr1}
{\cal D}^0 \,\mathscr{X} = 0\,,\qquad\left({\cal D}_z^{--} - {\cal D}^{--}\right) \mathscr{X} = 0\,,\qquad
\tilde{I}^0\,\mathscr{X} =
\tilde{F}\,\mathscr{X} =
\tilde{I}^{++}\,\mathscr{X} = 0\,,
\ee
and additional constraints
\begin{equation}  \label{x-constr2}
{\cal D}^{++} \,\mathscr{X}=0\,,
\end{equation}
\begin{equation}  \label{x-constr3}
{\cal D}^{-}{\cal D}^{+} \,\mathscr{X}=0\,,\qquad
\bar {\cal D}^{-}\bar {\cal D}^{+}\, \mathscr{X}=0\,,\qquad
\left({\cal D}^{-}\bar {\cal D}^{+} + \bar {\cal D}^{-}{\cal D}^{+}\right)\mathscr{X}=2m\mathscr{X}\,.
\end{equation}
The set of the constraints (\ref{x-constr1}) -- (\ref{x-constr3}) is invariant with respect
to ${\rm SU}(2|1)$ transformations. Indeed,
$\delta\left({\cal D}^{-}{\cal D}^{+} \,\mathscr{X} \right)=
-2m\left(\epsilon^-\bar\theta^++\bar\epsilon^-\theta^+ \right){\cal D}^{-}{\cal D}^{+} \,\mathscr{X}$, etc.
The constraints (\ref{x-constr1}) -- (\ref{x-constr3}) are solved by \footnote{
Note that
${\cal D}^{-}{\cal D}^{+} \,\mathscr{X}=\left(-\frac{\partial}{\partial \theta^+} - 2i \,\bar\theta^-\partial_{t_A}
-m\,\bar{\theta}^{-} \theta^-\frac{\partial}{\partial \theta^-}
+ m\,\bar\theta^+\frac{\partial}{\partial z^{++}} \right){\cal D}^{+} \,\mathscr{X}
-2m\,\bar\theta^-\,{\cal D}^{+} \,\mathscr{X}$, etc.,
because of (\ref{D-}) and (\ref{I0-op}).
}
\begin{equation}  \label{X-sol}
\begin{array}{rcl}
\mathscr{X}&=& x +\theta^- \psi^+ + \bar\theta^- \bar\psi^+ - \theta^+
\psi^- -
\bar\theta^+ \bar\psi^-
\\ [6pt]
&&
+\theta^-\bar\theta^- N^{++} + \theta^+\bar\theta^+ N^{--} + \theta^-\bar\theta^+ N -\theta^+\bar\theta^- \bar N
\\ [6pt]
&&
+ \,\theta^-\theta^+\bar\theta^- \Omega^+
+ \bar\theta^-\bar\theta^+\theta^- \bar\Omega^+
+ \theta^- \theta^+ \bar\theta^+ \Omega^-
+ \bar\theta^- \bar\theta^+ \theta^+ \bar\Omega^-
+ \theta^-\bar\theta^-\theta^+\bar\theta^+ D\,.
\end{array}
\end{equation}
Here,
\begin{equation}\label{N-+}
N^{\pm\pm} = N^{ik}w_i^\pm w_k^\pm \,,\quad  N = -i \partial_{t_A} x - N^{ik}w_i^+
w_k^- +m x \,,\quad  \bar N = i \partial_{t_A} x + N^{ik}w_i^+
w_k^- +m x\,,
\end{equation}
\begin{equation}\label{D-X}
D = 2\left( \partial_{t_A}\partial_{t_A}{x} +m^2 x -i \partial_{t_A}{N}{}^{ik}w_i^+ w_k^- \right)\,,
\end{equation}
\begin{equation}\label{psi-X}
\psi^{\pm} = \psi^{i}w_i^\pm  \,,\qquad \bar\psi{}^{\pm} =
\bar\psi{}^{i}w_i^\pm \,,
\qquad
\Omega^{-} = m\psi^{-}  \,,\qquad \bar\Omega^{-} =
m\bar\psi{}^{-}\,,
\end{equation}
\begin{equation}\label{om-X}
\Omega^{+} = -2i\partial_{t_A}{\psi}{}^+ -2m {\psi}{}^+ \,,\qquad \bar\Omega^{+} =
2i\partial_{t_A}{\bar\psi}{}^+ -2m{\bar\psi}{}^+
\end{equation}
and $x(t_A)$, $N^{ik}= N^{(ik)}(t_A)$, $\psi^{i}(t_A)$,
$\bar\psi_{i}(t)=(\overline{\psi^{i}})$ are $d{=}1$ fields.

After passing to the central basis coordinates by \p{var-new},
we observe that the $\theta$ expansion of the superfield (\ref{X-sol}) in the central basis takes the form \cite{IS14a}
\begin{equation}  \label{X-sol-c}
\begin{array}{rcl}
\mathscr{X}(t,\theta_i,\bar\theta^i)&=& x \,+ \theta_k\psi^k - \bar\theta^k\bar\psi_k
\, + m\,\theta_k\bar\theta^k \, x
+\theta^k\bar\theta^j N_{kj}
\\ [6pt]
&&
\displaystyle{
+\,\frac{1}{2}\,(\theta)^2\bar\theta^k\left(i\dot{\psi}_k+2m {\psi}_k\right)
-\frac{1}{2}\,(\bar\theta)^2\theta_k\left(i\dot{\bar\psi}{}^k-2m{\bar\psi}^k \right)
}
\\ [6pt]
&&
\displaystyle{
+\, (\theta)^2(\bar\theta)^2 \left(\frac{1}{4}\,\ddot{x} +m^2 x \right),
}
\end{array}
\end{equation}
where the component fields $x(t)$, $N^{ik}= N^{(ik)}(t)$, $\psi^{i}(t)$,
$\bar\psi_{i}(t)=(\overline{\psi^{i}})$ are the functions of real time $t$ and
$(\theta)^2\equiv \theta_i\theta^i$,
$(\bar\theta)^2\equiv \bar\theta^i\bar\theta_i$, $\dot x =\partial_{t_A}x$, {etc}.

The fermionic ${\rm SU}(2|1)$ transformations of component fields are the following
\be\label{143-tr}
\begin{array}{c}
\delta x =-\,\epsilon_k\psi^k +\bar{\epsilon}^k \bar\psi_k\,, \\ [6pt]
\delta \psi^k =i\,\bar{\epsilon}^k \dot{x} - \bar\epsilon_j N^{kj} -m\,\bar{\epsilon}^k x \,,\qquad
\delta \bar\psi_k =-i\, {\epsilon}_k \dot{x} - \epsilon^j N_{kj} -m\, {\epsilon}_k x\,, \\ [6pt]
\delta N^{kj} =-2i  \Big( \epsilon^{(k} \dot{\psi}^{j)} + \bar{\epsilon}^{(k} \dot{\bar\psi}^{j)} \Big)
-2m\Big( \epsilon^{(k} {\psi}^{j)} - \bar{\epsilon}^{(k} {\bar\psi}^{j)} \Big)\,.
\end{array}
\ee

The free $\mathscr{X}$-action reads
\be\lb{act-X}
S_{\mathscr{X}}=-\frac{1}{4}\,\int\,d\zeta_H\,\mathscr{X}^2\,.
\ee
Integrating in it over the $\theta$-variables
and harmonics \footnote{We use $ {\displaystyle\int} dw
\,w^{+i}w^-_k=\frac{1}{2}\,\delta^i_k $,\, $ {\displaystyle\int} dw \,w^{+(i_1}w^{+i_2)}w^-_{(k_1}
w^-_{k_2)}= -2{\displaystyle\int} dw \,w^{+(i_1}w^{-i_2)}w^+_{(k_1} w^-_{k_2)}=
\frac{1}{3}\,\delta^{(i_1}_{(k_1}\delta^{i_2)}_{k_2)} $.}, we obtain the component action \cite{IS14a}
\be\label{4N-X-WZ}
S_{\mathscr{X}} = \frac{1}{2}\,\displaystyle{\int} dt\,
\left[ \dot x\dot x +i \left(\bar\psi_k \dot\psi^k
-\dot{\bar\psi}_k \psi^k \right)
- m^2 x^2 + 2m\, \bar\psi_k \psi^k
- \frac{1}{2}\,N^{ik}N_{ik}\, \right] .
\ee

Another description of the multiplet ${\bf (1,4,3)}$ is through
an analytic real prepotential $\mathcal{V} (\zeta_A)$ (${\mathcal{D}}^{+} \,\mathcal{V}=\bar{\mathcal{D}}^{+}\, \mathcal{V}=0$). Its pregauge
freedom
\be \lb{ga-V0}
\delta \mathcal{V} = \mathcal{D}^{++}\lambda^{--}\,, \qquad \lambda^{--} =
\lambda^{--}(\zeta_A)\,,
\ee
can be exploited to show that $\mathcal{V}(\zeta_A)$ describes just the multiplet $({\bf 1, 4, 3})$
(by choosing the appropriate WZ gauge).
The superfield $\mathcal{V}(\zeta_A)$  is related to the superfield
$\mathscr{X}$ in the central basis by the harmonic integral transform
\begin{equation}  \label{X0-V0}
\mathscr{X}(t,\theta_i,\bar\theta^i)=\int dw\,
\Big(1+ m\,\theta^+\bar\theta^- -m\,\theta^-\bar\theta^+  \Big)^{-1}\,
\mathcal{V} \left(t_A,
\theta^+,
\bar\theta^+, w^\pm \right) \Big|\,,
\end{equation}
where the vertical bar $\Big|$ means that the expressions
$t_{A} = t + i\left(\theta^{+}\bar{\theta}^{-} + \theta^{-}\bar{\theta}^{+}\right)$,
$\theta^{-}=\theta^{i}w^{-}_{i}$, $\bar{\theta}^{-}= \bar{\theta}^{k} w^{-}_k$,
$\theta^{+}=
\theta^{i} w^{+}_{i} \left(1 + m\,\theta^{k} w^{-}_{k}\bar{\theta}^{l} w^{+}_l \right)$,
$\bar{\theta}^{+}=\bar{\theta}^{k} w^{+}_k
\left(1 - m\,\theta^{k} w^{+}_{k}\bar{\theta}^{l} w^{-}_l\right)$
defined in \p{var-new} should be substituted into the integrand.
Then, from (\ref{X0-V0}) we can identify the fields appearing in the WZ
gauge for $\mathcal{V}$ with the fields in \p{X-sol}
\begin{equation}  \label{V0-WZ}
\mathcal{V} (\zeta_A) =x(t_A)- 2\,\theta^+
\psi^{i}(t_A)w^-_i  -
2\,\bar\theta^+ \bar\psi^{i}(t_A)w^-_i + 3\,\theta^+ \bar\theta^+
N^{ik}(t_A)w^-_i
w^-_k
\,.
\end{equation}
The representation (\ref{X0-V0}) generalizes the analogous transform  in the ``flat'' non-deformed ${\cal N}{=}4$
supersymmetric mechanics \cite{DI06,DI07,FIL09,FIL10}.

The passive ${\rm SU}(2|1)$ transformation of the prepotential field $\mathcal{V}$ has the form
\be\label{trans-psi-sV}
\delta\mathcal{V} = -2 m
\left(\epsilon^-\bar\theta^+ + \bar\epsilon^-\theta^+\right) \mathcal{V} \,,
\ee
and the compensating gauge transformations for preserving the WZ gauge \p{V0-WZ} are
\be\label{trans-comp-V}
\delta_{comp}\mathcal{V} = \mathcal{D}^{++}\Lambda_{comp}^{--}\,,\quad
\Lambda_{comp}^{--}=
-\left(\epsilon^i\psi^j + \bar\epsilon^i \bar\psi^j\right)w^-_iw^-_j +
\left(\theta^+\bar\epsilon^i - \bar\theta^+\epsilon^i \right)N^{jk}w^-_iw^-_jw^-_k \,.
\ee
Applying \p{trans-psi-sV} and \p{trans-comp-V} to the WZ gauge expression \p{V0-WZ}, we reproduce the component field
transformations \p{143-tr}.

Note that \p{trans-psi-sV} agrees with the general transformation law \p{trans-psi} with
${\tilde I}^{++}\mathcal{V} =\tilde{F}\mathcal{V}= 0\,,$  $\tilde{I}^0\mathcal{V} = 2\,$
\footnote{The superfield $\mathcal{V}$ supplies an example of analytic ${\rm SU}(2|1)$ superfield not satisfying the constraint \p{I0-cond}. This property is harmless
because $\mathcal{V}$ is not subject to any extra harmonic constraints. One can formally define ${\cal D}^{++}\mathcal{V}$, and it is a covariant ${\rm SU}(2|1)$
analytic superfield living on the superspace $\hat{\zeta}_A$ \p{AnalSubs} and having a linear dependence on $z^{++}$ (in the $(w^\pm_i, z^{++})$ basis).}.
Using the transformation of the harmonic measure $\delta \,dw = \partial_w^{--}(\eta^+\bar\theta^+ + \bar\eta^+\theta^+)\, dw$ in the central basis, it is straightforward to be
convinced that \p{trans-psi-sV} just reproduces the transformation $\delta \mathscr{X} = 0\,$ for $\mathscr{X}$ defined in \p{X0-V0}.

\subsection{The multiplet ${\bf (4,4,0)}$ and ${\rm SU}(2|1)$ invariant WZ term}

The multiplet ${\bf (4,4,0)}$ is described by the superfield
$\mathscr{Z}^{+} (t_A, \theta^\pm, \bar\theta^\pm, z^{++}, u^\pm)$
possessing the unit ${\rm U}(1)$ charge,
\be\label{z-constr0}
{\cal D}^0 \,\mathscr{Z}^{+} = \mathscr{Z}^{+}\,,
\ee
and satisfying the ${\rm SU}(2|1)$ covariant constraints
\be\label{z-constr1}
\left({\cal D}_z^{--} - {\cal D}^{--}\right) \mathscr{Z}^{+} = 0\,,\qquad
\tilde{I}^0\,\mathscr{Z}^{+} =-\mathscr{Z}^{+}\,,\qquad
\tilde{F}\,\mathscr{Z}^{+} =
\tilde{I}^{++}\,\mathscr{Z}^{+} = 0\,,
\ee
as well as
\begin{equation}  \label{z-constr2}
{\cal D}^{++} \,\mathscr{Z}^{+}=0\,,
\end{equation}
\begin{equation}  \label{z-constr3}
{\cal D}^{+} \mathscr{Z}^{+} =\bar{\cal D}^{+} \mathscr{Z}^{+} = 0\,.
\end{equation}

The constraints \eqref{z-constr3} together with $\tilde{I}^{++}\,\mathscr{Z}^{+} = 0$ imply
the superfield $\mathscr{Z}^{+}$ to be analytic,
that is
\be\label{z-sol1}
\mathscr{Z}^{+}(t_A, \theta^+, \bar\theta^+, u^\pm, z^{++}) = {\mathcal{Z}}^{+} (t_A, \theta^+, \bar\theta^+, w^\pm) = {\mathcal{Z}}^{+}(\zeta_A)\,.
\ee
The general solution of the full set of the constraints \eqref{z-constr0} -- \eqref{z-constr3}
is represented by the component expansion of the harmonic superfield \eqref{z-sol1}
in the following form \cite{IS15}
\be\label{z-sol2}
{\mathcal{Z}}^{+} (t_A, \theta^+, \bar\theta^+, w^\pm) = z^{i}w^{+}_i + \theta^{+} \varphi +
\bar\theta^{+} \phi -
2i \theta^{+}\bar{\theta}^{+} \dot{Z}^{i} w^{-}_i\,.
\ee

The fermionic ${\rm SU}(2|1)$ transformation of $\mathscr{Z}^{+}$ is a particular case of the general transformation law \eqref{trans-psi-s},
\be\label{440-tr-f}
\delta \mathscr{Z}^{+} = m\left(\epsilon^{-}\bar{\theta}^{+} + \bar\epsilon^{-}\theta^{+}\right)\mathscr{Z}^{+}\,.
\ee
It implies the following transformations for the component fields
\be\label{440-tr}
\begin{array}{lll}
\delta z^{i} =-\,\epsilon^i\varphi -\bar{\epsilon}^i \phi \,,\quad &
\delta \varphi =2i\bar{\epsilon}^k \dot{z}_k+ m\,\bar{\epsilon}^k z_k \,,\quad &
\delta \phi =2i\epsilon_k \dot{z}^{k} - m\,\epsilon_k z^{k}\,, \\ [6pt]
\delta \bar z_{i} = \epsilon_i\bar\phi -\bar{\epsilon}_i \bar\varphi \,,\quad &
\delta \bar\varphi =2i\epsilon_k \dot{\bar z}^{k} - m\,\epsilon_k \bar z^{k}\,,\quad &
\delta \bar\phi =-\,2i\bar{\epsilon}^k \dot{\bar z}_k- m\,\bar{\epsilon}^k \bar z_k \,.
\end{array}
\ee

It has been shown in  \cite{IS15} that the Wess-Zumino type actions enjoying ${\rm SU}(2|1)$ supersymmetry cannot
be constructed for the single multiplet ${\bf (4,4,0)}$.
However, if we couple the multiplet ${\bf (4,4,0)}$ \eqref{z-sol1} to the multiplet ${\bf (1,4,3)}$ \p{X-sol}, \p{V0-WZ}
the ${\rm SU}(2|1)$-invariant WZ action can be set up.

Such WZ action is given by the following integral over the analytic subspace
\be\label{act-WZ}
S_{\rm WZ}({\mathcal{V}},{\mathcal{Z}}^{+}) =\frac{1}{2} \int d\zeta^{--}_A \,{\mathcal{V}}\,{\mathcal{Z}}^{+}\tilde{\mathcal{Z}}^{+}\,,
\ee
where $\tilde{\mathcal{Z}}^{+}$ is generalized harmonic conjugate of ${\mathcal{Z}}^{+}$ (see \cite{HSS,IS15} for definition of such conjugation).
As a consequence of \eqref{anmeas}, \p{trans-psi-sV} and \p{440-tr-f}, the action \p{act-WZ} is ${\rm SU}(2|1)$ invariant.
The corresponding component action $S_{\rm WZ}={\displaystyle\int} dt\, L_{\rm WZ}$ with the component Lagrangian
\be\label{lagr-WZ}
\begin{array}{rcl}
L_{\rm WZ} & = &{\displaystyle -\frac{i}{2}\,x\Big( {\bar z}_{k}\dot{z}^{k}-\dot{\bar z}_{k}{z}^{k}\Big)
- \frac{1}{2} \, {N}^{kj} {z}_{k}{\bar z}_{j}} \\ [8pt]
&& {\displaystyle + \frac{1}{2} \,\psi^k\Big( {z}_{k}\bar\varphi +{\bar z}_{k}\phi\Big)
+ \frac{1}{2} \,\bar\psi^k\Big( {z}_{k}\bar\phi -{\bar z}_{k}\varphi\Big)
+ \frac{1}{2} \,x\Big( \varphi\bar\varphi +\phi\bar\phi\Big)}
\end{array}
\ee
is invariant under the ${\rm SU}(2|1)$ transformations \eqref{143-tr}, \eqref{440-tr}.

\subsection{Total action}

Now we consider a system with the action given by the sum
$S_{\mathscr{X}} + S_{\rm WZ}$. Making use of the component form of these actions defined in
\eqref{4N-X-WZ} and \eqref{lagr-WZ}, eliminating the auxiliary fields $\phi$, $\bar\phi$, $\varphi$, $\bar\varphi$, $N^{ik}$
from this sum by their algebraic equations of motion
\be\label{aux-fields}
N^{ik}=-z^{(i}\bar z^{k)}\,,\qquad\varphi=-\psi^k z_k/x\,,\quad \bar\varphi=-\bar\psi^k \bar z_k/x\,,\qquad
\phi=-\bar\psi^k z_k/x\,,\quad \bar\phi=\psi^k \bar z_k/x
\ee
and, finally, redefining   ${z}^{k}\to {z}^{k}/\sqrt{x}\,$, we obtain
\be\label{total-X-WZ}
\begin{array}{rcl}
S_{\mathscr{X}} +S_{\rm WZ}&=& \displaystyle{\int dt\,
\left\{ \frac{1}{2}\,\dot x\dot x +\frac{i}{2}\,\left(\bar\psi_k \dot\psi^k
-\dot{\bar\psi}_k \psi^k \right)\right.
-\frac{i}{2}\,\Big( {\bar z}_{k}\dot{z}^{k}-\dot{\bar z}_{k}{z}^{k}\Big)\,  }\\ [8pt]
&&\qquad\qquad \displaystyle{\left.
- \frac{1}{2}\,m^2 x^2 + m\, \bar\psi_k \psi^k-\frac{1}{x^2}\,\left[ \frac{1}{8}\,({z}^{k}\bar z_{k})^2 +
\psi^i\bar\psi^k {z}_{(i}\bar z_{k)} \right]\right\}.}
\end{array}
\ee
In contrast to the analogical model  of the ${\cal N}=4$ supersymmetric mechanics \cite{FIL09,FIL10},
the action (\ref{total-X-WZ}) contains mass term (oscillator term) for the component field $x$.
But the spinning variables ${z}^{i}$ prove to be not restricted by any constraint besides the second class constraints
produced by the first order kinetic term for these variables. As a result, the quantum spectrum of this
composite model involves an infinite number of the states, like in its ``flat'' prototype.

For getting the finite number of physical states it is necessary to impose an additional constraint
which amounts to the gauging procedure described in the next section.

\setcounter{equation}0
\section{Gauging of coupled dynamical multiplet ${\bf (1,4,3)}$ and \\ semi-dynamical multiplet ${\bf (4,4,0)}$}

The WZ action \eqref{act-WZ} and the total action $S_{\mathscr{X}} + S_{\rm WZ}$ are invariant with respect to the global ${\rm U}(1)$
transformations
\be\label{tran4-Phi}
\mathscr{Z}^+{}' = e^{i\lambda} \mathscr{Z}^+,\qquad
\tilde{\mathscr{Z}}{}^+{}' =
e^{-i\lambda}\tilde{\mathscr{Z}}{}^+\,.
\ee
Now we require local invariance of this action, with the parameter in (\ref{tran4-Phi})
being promoted to an analytic superfield $\lambda = \lambda(\zeta_A)$ satisfying the conditions
\be\label{l-constr1}
{\cal D}^{+} \lambda =\bar{\cal D}^{+} \lambda = 0\,,
\qquad \left({\cal D}_z^{--} - {\cal D}^{--}\right) \lambda = 0\,,
\qquad
{\cal D}^{0} \lambda =\tilde{I}^0\lambda =\tilde{F}\lambda =\tilde{I}^{++}\lambda = 0\,.
\ee

To secure this local symmetry in the considered system we introduce the Grassmann-even analytic
gauge superfield $V^{++}$, which satisfies the conditions
\be\label{V-constr1}
{\cal D}^{+} V^{++} =\bar{\cal D}^{+} V^{++} = 0\,,
\qquad \tilde{I}^{++}V^{++} = 0\,,
\ee
\be\label{V-constr2}
\left({\cal D}_z^{--} - {\cal D}^{--}\right) V^{++} = 0\,,
\qquad {\cal D}^{0} V^{++} =-\tilde{I}^0V^{++} = 2V^{++}\,,\qquad
\tilde{F}V^{++} = 0
\ee
and is defined up to the gauge transformations
\be\label{tran4-V}
V^{++}{}' = V^{++} - D^{++}\lambda\,.
\ee

The gauge superfield $V^{++}$ covariantizes the derivative $\mathscr{D}^{++}$.
As a result, the complex analytic superfield ${\mathcal{Z}}^+,
\tilde{\mathcal{Z}}^+$, instead of the constraints (\ref{z-constr3}), gets subjected to the covariantized harmonic constraints
\begin{equation}  \label{cons-Ph-g}
\nabla^{++} \,{\mathcal{Z}}^+\equiv (\mathscr{D}^{++} + i\,V^{++})
\,{\mathcal{Z}}^+=0\,,\qquad \nabla^{++}
\,\tilde{\mathcal{Z}}{}^+\equiv (\mathscr{D}^{++} - i\,V^{++}) \,\tilde{\mathcal{Z}}{}^+=0\,.
\end{equation}

We can also add to the total action the gauge-invariant Fayet-Iliopoulos (FI) term
\begin{equation}\label{4N-FI-1}
S_{FI} = \frac{i}{2}\,\,c\int \mu^{(-2)}_A \,V^{++}\,.
\end{equation}
So, we will consider the action
\begin{equation}\label{4N-gau-1}
S =S_{\mathscr{X}} + S_{WZ} + S_{FI}\,.
\end{equation}

Using the ${\rm U}(1)$ gauge freedom~\p{tran4-V}, (\ref{tran4-Phi}) we can
choose the WZ gauge
\begin{equation}  \label{WZ-4Na}
V^{++} =2i\,\theta^{+} \bar\theta^{+}A(t_A)\,.
\end{equation}
Then
\begin{equation}  \label{4N-FI-WZ}
S_{FI} = - c \int dt \,A\,.
\end{equation}

The solution of the constraint~(\ref{cons-Ph-g}) in the WZ gauge~(\ref{WZ-4Na}) is
\begin{equation}  \label{Ph-WZ0}
\begin{array}{rcl}
\mathscr{Z}^+(t_A, \theta^+, \bar\theta^+, u^\pm, z^{++}) &=& z^{i}w_i^+ + \theta^+ \varphi + \bar\theta^+ \phi - 2i\,
\theta^+
\bar\theta^+\nabla_{t_A}z^{i}w_i^-
\,,\\ [7pt]
\tilde{\mathscr{Z}}{}^+(t_A, \theta^+, \bar\theta^+, u^\pm, z^{++})  &=& \bar z_{i}w^{+i} + \theta^+ \bar\phi -
\bar\theta^+
\bar\varphi - 2i\, \theta^+ \bar\theta^+ \nabla_{t_A}\bar z_{i}w^{-i}\,,
\end{array}
\end{equation}
where
\begin{equation}\label{cov-Z}
\nabla z^k=\dot z^k + i A \, z^k\,, \qquad\nabla \bar z_k=\dot{\bar z}_k
-i A\,\bar
z_k\,.
\end{equation}

Plugging the expressions (\ref{Ph-WZ0}) and (\ref{V0-WZ}) into the
action (\ref{act-WZ}) and integrating there over $\theta\,$s and harmonics, we
obtain the component form of the WZ action
\be\label{4N-WZ-WZ}
\begin{array}{rcl}
S_{WZ} &=& {\displaystyle -\frac{i}{2}\int dt \,\Big(\bar z_k \nabla z^k -
\nabla \bar z_k \, z^k \Big)x - \frac{1}{2}\int dt \,N^{ik}\bar z_i z_k }
\\ [9pt]
&& {\displaystyle  \, + \frac{1}{2}\int dt \, \Big[\psi^k
\Big(\bar\varphi\,z_k+
\bar z_k\phi\Big) + \bar\psi^k \Big(\bar{\phi\,\,}\! z_k- \bar
z_k\varphi\Big) - x
\Big(\bar{\phi\,\,}\! \phi+ \bar\varphi\,\varphi\Big)\Big] \, .}
\end{array}
\ee
The fermionic fields $\phi, \varphi$ are auxiliary. The action is
invariant under
the residual local ${\rm U}(1)$ transformations
\be
A' = A - \dot{\lambda}_0\,, \qquad z^i{}' = e^{i\lambda_0}z^i\,, \quad
\bar{z}_i{}' =
e^{-i\lambda_0}\bar{z}_i \label{res-ga}
\ee
(and similar phase transformations of the fermionic fields).

The total  action (\ref{4N-gau-1}) in the WZ gauge takes the following on-shell form (like in (\ref{total-X-WZ}), we
make the redefinition ${z}^{k}\to {z}^{k}/\sqrt{x}$)
\begin{eqnarray}
S &=& S_b+ S_f\,, \label{4N-ph}\\
S_b &=&  \frac12 \int dt \,\left[\dot x\dot x  -m^2 x^2 + i \left(\dot{\bar z}_k z^k -
{\bar z}_k \dot z^k\right)-\frac{(\bar z_k
z^{k})^2}{4x^2} +2A \left(\bar z_k z^{k} -c \right) \right] \,,\label{bose}\\
S_f &=&  \int  dt \Big[ \frac{i}{2}\Big( \bar\psi_k \dot\psi^k -\dot{\bar\psi}_k \psi^k \Big) + m\bar\psi_k\psi^k\Big] -
\int  dt \, \frac{\psi^{i}\bar\psi^{k} z_{(i} \bar z_{k)}}{x^2} \,. \label{fermi}
\end{eqnarray}
The last term in the bosonic action (\ref{bose}) produces first class constraint $\bar z_k z^{k} -c \approx 0$
restricting the quantum spectrum to a single supermultiplet.

\setcounter{equation}0
\section{Matrix model}

Now we are going to generalize the model of the previous section to the ${\rm U}(n)$, $d{=}1$ gauge theory
following the papers \cite{FIL08,FIL12}.

The matrix model to be constructed involves the following ${\rm U}(n)$ entities:
\begin{itemize}
\item
$n^2$ commuting superfields $\mathscr{X}_b^a=(\widetilde{\mathscr{X}_a^b})$, $a,b=1,\ldots ,n$
forming the  hermitian $n{\times}n$-matrix superfield
$ \mathscr{X}=(\mathscr{X}_a^b)$ in adjoint representation of ${\rm U}(n)$;
\item
$n$ commuting complex superfields ${\mathscr{Z}}{}^{+}_a$ forming the ${\rm U}(n)$ spinor $\mathscr{Z}^+=(\mathscr{Z}^+_a)$,
$\tilde{\mathscr{Z}}^{+}=(\tilde {\mathscr{Z}}^{+a})$;
\item
$n^2$ non-propagating ``gauge superfields'' $ V^{++}=(V^{++}{}_a^b)$, $(\widetilde{V^{++}{}_a^b}) =V^{++}{}_b^a$.
\end{itemize}
The local ${\rm U}(n)$ transformations are given by
\begin{equation}\label{tran4}
\mathscr{X}^{\,\prime} =  e^{i\lambda} \mathscr{X} e^{-i\lambda} , \qquad
\mathcal{Z}^+{}^{\prime}
= e^{i\lambda} \mathcal{Z}^+ , \qquad
V^{++}{}^{\,\prime} =  e^{i\lambda}\, V^{++}\, e^{-i\lambda} - i\, e^{i\lambda} (D^{++}
e^{-i\lambda}),
\end{equation}
where $ \lambda_a^b(\zeta_A) \in u(n) $ is the ``hermitian'' analytic matrix
parameter, $\widetilde{\lambda} =\lambda$.

The ${\rm SU}(2|1)$ supersymmetric matrix model with ${\rm U}(n)$ gauge symmetry is described by
the action
\begin{equation}\label{4N-gau}
\mathscr{S}_{matrix} =\mathscr{S}_{\mathscr{X}} + \mathscr{S}_{WZ} + \mathscr{S}_{FI}\,.
\end{equation}

The first term in (\ref{4N-gau}),
\begin{equation}\label{4N-X}
\mathscr{S}_{\mathscr{X}} =-\frac{1}{4}\int \mu_H  {\rm Tr} \left( \mathscr{X}^{\,2} \,
\right),
\end{equation}
is the gauged action of the ${\bf (1, 4, 3)}$ multiplets. Now the superfields $\mathscr{X}=(\mathscr{X}_a^b)$ are
subjected to the constraints (\ref{x-constr1}) and
\begin{equation}  \label{x-constr-cov2}
\nabla^{++} \,\mathscr{X}= \mathcal{D}^{++}
\mathscr{X} + i\,[V^{++} ,\mathscr{X}]=0\,,
\end{equation}
\begin{equation}  \label{x-constr-cov3}
\nabla^{-}\nabla^{+} \,\mathscr{X}=0\,,\qquad
\bar \nabla^{-}\bar \nabla^{+}\, \mathscr{X}=0\,,\qquad
\left(\nabla^{-}\bar \nabla^{+} + \bar \nabla^{-}\nabla^{+}\right)\mathscr{X}=2m\mathscr{X}\,,
\end{equation}
which are gauge-covariantization of the constraints (\ref{x-constr2}), (\ref{x-constr3}).
The constraint (\ref{x-constr-cov2}) involves the covariant harmonic derivative
$\nabla^{++} = \mathcal{D}^{++} + i\,V^{++}$, where the gauge matrix connection
$V^{++}(\zeta,w)$ is an analytic superfield.\footnote{Besides the covariant derivative
$\nabla^{++}$ which commutes with $\mathcal{D}^+, \bar{\mathcal{D}}^+$ and so preserves the analyticity,
one can define the derivative $\nabla^{--} = \mathcal{D}^{--} + i\,V^{--}$, so that
$[\nabla^{++}, \nabla^{--}] = \mathcal{D}^0\,$. The non-analytic connection $V^{--}$ is expressed
through $V^{++}$ from this commutation relation \cite{HSS}.} The gauge connections
entering the spinor covariant derivatives in~(\ref{x-constr-cov3}) are properly expressed
through $V^{++}(\zeta,u)$. The parameters of the ${\rm U}(n)$ gauge group are
analytic, so $\nabla^{+} = \mathcal{D}^{+}\,,\;\bar{\nabla}^{+} = \bar{\mathcal{D}}^{+}$.

The last term in~(\ref{4N-gau}) is the FI term
\begin{equation}\label{4N-FI}
\mathscr{S}_{FI} = \frac{i}{2}\,c\int \mu^{(-2)}_A  \,{\rm Tr} \,V^{++}\,,
\end{equation}
whereas the second term,
\begin{equation}\label{4N-VZ}
\mathscr{S}_{WZ} = \frac{1}{2}\int \mu^{(-2)}_A  \mathcal{V}_0
\widetilde{\,\mathcal{Z}}{}^{+a} \mathcal{Z}^+_a\,,
\end{equation}
is a WZ action describing coupling of $n$ commuting analytic superfields $\mathcal{Z}^+_a$ and
the singlet ${\rm U}(1)$ part $\mathscr{X}_0 \equiv {\rm Tr} \left( \mathscr{X} \right)$.
The real analytic superfield $\mathcal{V}_0(\zeta,w)$ is defined by the integral
transform (\ref{X0-V0}) for the trace part:
\begin{equation}  \label{X0-V0-tr}
\mathscr{X}_0(t,\theta_i,\bar\theta^i)=\int dw
\Big(1+m\theta^-\bar\theta^+ - m\theta^+\bar\theta^- -2m^2\theta^+\theta^-\bar\theta^+\bar\theta^-\Big)
\mathcal{V}_0 \left(t_A,
\theta^+,
\bar\theta^+, w^\pm \right) \Big|\,.
\end{equation}
The $n$ multiplets ${\bf (4,4,0)}$ are described by the superfields $\mathcal{Z}^+_a$
defined by the constraints (\ref{z-constr1}) -- (\ref{z-constr3})
in which the constraint ${\cal D}^{++} \mathscr{Z}^{+} = 0$
is gauge-covariantized:
\be\label{z-constr3-cov}
\nabla^{++} \mathscr{Z}^{+}=\left(\mathcal{D}^{++} + iV^{++}\right) \mathscr{Z}^{+} = 0\,.
\ee

Using the gauge freedom (\ref{tran4}) we can choose the WZ
gauge
\begin{equation}  \label{WZ-4N}
V^{++} =2i\,\theta^{+}
  \bar\theta^{+}A(t_A) \,,
\end{equation}
where now $A(t_A)$ is an $n{\times}n$ matrix field. In this gauge we have
\begin{equation}  \label{WZ-4N-c}
\nabla^{\pm\pm}= \mathcal{D}^{\pm\pm} -2\,\theta^{\pm}  \bar\theta^{\pm} A, \qquad
\nabla^{-}=\mathcal{D}^{-} +2\,  \bar\theta^{-} A, \qquad \bar{\nabla}^{-}= \bar{\mathcal{D}}^{-} +2\,
  \theta^{-} A\,.
\end{equation}
The solution to the constraints (\ref{x-constr1})
and the constraints (\ref{x-constr-cov2}), (\ref{x-constr-cov3})
for matrix field $\mathscr{X}$ is similar to (\ref{x-constr-cov3}) and
it is as follows:
\begin{equation}  \label{X-sol-M}
\begin{array}{rcl}
\mathscr{X}&=& X +\theta^- \Psi^+ + \bar\theta^- \bar\Psi^+ - \theta^+
\Psi^- -
\bar\theta^+ \bar\Psi^-
\\ [6pt]
&&
+\theta^-\bar\theta^- N^{++} + \theta^+\bar\theta^+ N^{--} + \theta^-\bar\theta^+ N -\theta^+\bar\theta^- \bar N
\\ [6pt]
&&
+ \,\theta^-\theta^+\bar\theta^- \Omega^+
+ \bar\theta^-\bar\theta^+\theta^- \bar\Omega^+
+ \theta^- \theta^+ \bar\theta^+ \Omega^-
+ \bar\theta^- \bar\theta^+ \theta^+ \bar\Omega^-
+ \theta^-\bar\theta^-\theta^+\bar\theta^+ D\,.
\end{array}
\end{equation}
Here,
\begin{equation}\label{N-+-M}
N^{\pm\pm} = N^{ik}w_i^\pm w_k^\pm \,,\quad  N = -i \nabla_{t_A} X - N^{ik}w_i^+
w_k^- +m X \,,\quad  \bar N = i \nabla_{t_A} X + N^{ik}w_i^+
w_k^- +m X\,,
\end{equation}
\begin{equation}\label{D-X-M}
D = 2\left( \nabla_{t_A}\nabla_{t_A}{X} +m^2 x -i \nabla_{t_A}{N}{}^{ik}w_i^+ w_k^- \right)\,,
\end{equation}
\begin{equation}\label{psi-X-M}
\Psi^{\pm} = \Psi^{i}w_i^\pm  \,,\qquad \bar\Psi{}^{\pm} =
\bar\Psi{}^{i}w_i^\pm \,,
\qquad
\Omega^{-} = m\Psi^{-}  \,,\qquad \bar\Omega^{-} =
m\bar\Psi{}^{-}\,,
\end{equation}
\begin{equation}\label{om-X-M}
\Omega^{+} = -2i\nabla_{t_A}{\Psi}{}^+ -2m {\psi}{}^+ \,,\qquad \bar\Omega^{+} =
2i\nabla_{t_A}{\bar\Psi}{}^+ -2m{\bar\Psi}{}^+\,.
\end{equation}
The quantities $X(t_A)$, $N^{ik}= N^{(ik)}(t_A)$, $\Psi^{i}(t_A)$,
$\bar\Psi_{i}(t_A)=({\Psi^{i}})^\dagger$ in (\ref{N-+-M}) -- (\ref{om-X-M}) are matrix $d{=}1$ fields
and the covariant
derivatives are defined by
\begin{equation}\label{cov-der-b-M}
\begin{array}{ll}
\nabla_{t_A}X =
\partial_{t_A}X +i[ A,X] \,,\qquad & \nabla_{t_A}N^{ik} =
\partial_{t_A}N^{ik} +i[ A,N^{ik}]\,, \\ [7pt]
\nabla_{t_A}\Psi^{i} =
\partial_{t_A}\Psi^{i} +i[ A,\Psi^{i}] \,,\qquad & \nabla_{t_A}\bar\Psi_{i} =
\partial_{t_A}\bar\Psi_{i} +i[ A,\bar\Psi_{i}]\,.
\end{array}
\end{equation}
The solution of the constraints (\ref{z-constr1}) -- (\ref{z-constr3}) with the covariantization (\ref{z-constr3-cov})
for ${\rm U}(n)$ spinor superfield $\mathscr{Z}^+$ is similar to (\ref{Ph-WZ0}):
\begin{equation}  \label{Ph-WZ0-M}
\begin{array}{rcl}
\mathscr{Z}^+(t_A, \theta^+, \bar\theta^+, u^\pm, z^{++}) &=& Z^{i}w_i^+ + \theta^+ \varphi + \bar\theta^+ \phi - 2i\,
\theta^+
\bar\theta^+\nabla_{t_A}Z^{i}w_i^-
\,,\\ [7pt]
\tilde{\mathscr{Z}}{}^+(t_A, \theta^+, \bar\theta^+, u^\pm, z^{++})  &=& \bar Z_{i}w^{+i} + \theta^+ \bar\phi -
\bar\theta^+
\bar\varphi - 2i\, \theta^+ \bar\theta^+ \nabla_{t_A}\bar Z_{i}w^{-i}\,,
\end{array}
\end{equation}
where
\begin{equation}\label{cov-Z-M}
\nabla Z^k=\dot Z^k + i A \, Z^k\,, \qquad\nabla \bar Z_k=\dot{\bar Z}_k -i A\,\bar Z_k
\end{equation}
are covariant derivatives of ${\rm U}(n)$ spinor $d{=}1$ fields $Z^{i}_a$, $\bar Z_{i}^a = (\overline{Z^{i}_a})$.

Inserting the expressions~(\ref{X-sol-M}), (\ref{Ph-WZ0-M}) in the action (\ref{4N-gau}) and
eliminating the fields $N^{ik}$, $\phi$,  $\bar\phi$, $\varphi$,  $\bar\varphi$ by their
equations of motion we obtain, in the WZ gauge,
\begin{eqnarray}\label{4N-gau-bose-a}
\mathscr{S}_{matrix}  &=& \mathscr{S}_b + \mathscr{S}_f,
\\
\mathscr{S}_b &=& \frac12\, {\rm Tr}\int dt \,\Big( \nabla X\nabla X - m^2 X^2 \Big)-c\int dt \,{\rm Tr}A
\nonumber \\
&& + \, \frac12\, {\rm Tr}\int dt \,
\Big[\,i X_0 \left(\nabla \bar Z_k \, Z^k-\bar Z_k \nabla Z^k \right)- \frac{n}{4}\,(\bar Z^{(i} Z^{k)})(\bar Z_{i} Z_{k})
\Big],\label{4N-gau-bose-1}
\\
\mathscr{S}_f&=&   \frac12\,{\rm Tr} \int dt \Big[i\left( \bar\Psi_k \nabla\Psi^k
-\nabla\bar\Psi_k \Psi^k
\right) +2m \bar\Psi_k \Psi^k\Big]  -\int dt  \,\frac{\Psi^{(i}_0\bar\Psi^{k)}_0 (\bar Z_{i}Z_{k})}{X_0}\,,\label{4N-gau-fermi-1}
\end{eqnarray}
where
$$
X_0 \equiv {\rm Tr} (X), \qquad\Psi_0^i \equiv {\rm Tr} (\Psi^i), \qquad\bar\Psi_0^i \equiv
{\rm Tr} (\bar\Psi^i)
$$
and $(\bar Z_{i}Z_{k})\equiv \bar Z_{i}^a Z_{ka}$, $(\nabla \bar Z_k \, Z^k)\equiv \nabla \bar Z_k^a \, Z^k_a$.

Let us consider the bosonic limit of $S_{matrix}$, {\it i.e.} the action (\ref{4N-gau-bose-1}).
Using the residual gauge invariance of the action (\ref{4N-gau-bose-1}),
$ X^{\,\prime} =  e^{i\lambda}\, X\, e^{-i\lambda} $,
$Z^{\prime}{}^{k} =  e^{i\lambda} Z^{k}$, $ A^{\,\prime} =  e^{i\lambda}\, A\,
e^{-i\lambda} - i\, e^{i\lambda} (\partial_t e^{-i\lambda})\,$, where $ \lambda_a^b(t) \in
u(n) $ are ordinary $d{=}1$  gauge parameters, we can
impose the gauge
$$
X_a^b =0\,, \qquad a\neq b\,,
$$
{\it i.e.} $X_a^b =X_a \delta_a^b$ and $X_0={\displaystyle \sum_{a=1}^n }X_a$.
As a result of this, and after eliminating
$A_a^b$, $a\neq b$, by the equations of motion, the action~(\ref{4N-gau-bose-1}) takes the
following form (instead of $Z^i_a$ we introduce the new fields $ Z^\prime{}^i_a =
(X_0)^{1/2}\,Z^i_a$ and omit the primes on these fields),
\begin{eqnarray}
\mathscr{S}_{b} &=& \frac12 \int dt \Bigg\{ \sum_{a} \Big(\dot X_a \dot X_a - m^2 X_a X_a\Big) - \frac{i}{2}\sum_{a} \Big(\bar
Z_k^a \dot Z^k_a - \dot {\bar Z}{}_k^a Z^k_a\Big)
+ 2\sum_{a} A_a^a\Big(Z_k^a Z^k_a -c\Big)
+ \nonumber\\
&& \qquad\qquad  + \sum_{a\neq b} \, \frac{{\rm Tr}(S_a S_b)}{4(X_a - X_b)^2} - \frac{n\,{\rm
Tr}(\hat S \hat S)}{2(X_0)^2}\,\Bigg\}, \label{4N-bose-fix}
\end{eqnarray}
where we used the following notation:
\begin{eqnarray}\label{S}
(S_a)_k{}^j &\equiv& \bar Z^a_k Z_a^j,\\ [6pt]
(\hat S)_k{}^j &\equiv& \sum_a \left[ (S_a)_k{}^j -
{\textstyle\frac{1}{2}}\delta_k^j(S_a)_l{}^l\right] \label{hS}
\end{eqnarray}
and no sum over the repeated index $a$ in (\ref{S}) is assumed.

The terms ${\displaystyle\sum_{a}} A_a^a\left(Z_k^a Z^k_a -c\right)$ in (\ref{4N-bose-fix})
produce $n$ constraints (for each index $a$)
\begin{equation}\label{4N-eq-aa}
\bar Z_k^a Z^k_a -c \approx 0
\end{equation}
for the fields $Z^k_a$.
The constraints (\ref{4N-eq-aa}) generate abelian gauge $[{\rm U}(1)]^n$ symmetry, $Z_a^k \rightarrow
e^{i\varphi_a} Z_a^k\,$, with local parameters $\varphi_a(t)$.

Due to the constraints (\ref{4N-eq-aa}), the fields $Z^k_a$
describe $n$ sets of the target harmonics. After quantization, these variables become purely internal
(${\rm U}(2)$-spin) degrees of freedom.
So, in the Hamiltonian approach, the kinetic WZ term for $Z$ in~(\ref{4N-bose-fix}) gives rise
to the following Dirac brackets:
\begin{equation}\label{DB}
[ Z_a^k, \bar Z^b_j]_{{}_D}= -i\delta^b_a\delta_j^k.
\end{equation}
With respect to these brackets the quantities~(\ref{S}) for each index $a$ form $u(2)$
algebras
\begin{equation}\label{su-DB}
[(S_a)_i{}^j, (S_b)_k{}^l]_{{}_D}= i\delta_{ab}\left\{\delta_i^l(S_a)_k{}^j-\delta_k^j(S_a)_i{}^l
 \right\}.
\end{equation}
As a result, after quantization the variables $Z^k_a$ describe $n$ sets of fuzzy spheres.

The action (\ref{4N-bose-fix}) contains
a potential in the center-of-mass sector with the coordinate $X_0$  (last term in~(\ref{4N-bose-fix})). Modulo
this extra potential, the bosonic limit of the
system constructed is none other than the U(2)-spin Calogero-Moser
model which is a massive generalization of the U(2)-spin Calogero
model \cite{GiHe,Woj} in the formulation of \cite{Poly}.

\setcounter{equation}0
\section{Concluding remarks and outlook}

In this paper, we proposed new models of ${\rm SU}(2|1)$ supersymmetric quantum mechanics as
a deformation  of the corresponding ``flat'' ${\cal N}=4, d=1$ supersymmetric models. The characteristic
features of these models is the use of different types of
supermultiplets: dynamical, semi-dynamical and pure gauge ones. In
considered models, dynamical multiplets are the ${\bf (1,4,3)}$ ones. The prepotential superfield  description of them has
provided an opportunity to build the WZ action for the  ${\bf
(4,4,0)}$ multiplets and thereby to use the latter for describing semi-dynamical degrees of freedom. The ${\rm SU}(2|1)$ version
of the superfield gauging procedure of refs. \cite{DI06,DI07} involving the appropriate gauge multiplets allowed us to gauge away
some of the dynamical and semi-dynamical fields on shell.

We have studied these new  ${\rm SU}(2|1)$ supersymmetric mechanics models both in the one-particle case and in the multi-particle one.
In the latter case the system is described off shell by the matrix theory
with ${\rm U}(n)$ gauging. After elimination of auxiliary and pure gauge fields
this matrix theory yields new ${\cal N}=4$ superextensions of the
$A_{n-1}$ Calogero-Moser model. The mass (frequency) of the physical states is defined
by the deformation parameter of the ${\rm SU}(2|1)$ supersymmetry.

The ${\cal N}=4$ superextensions of the Calogero-Moser model play a crucial role
in applying the multiparticle integrable Calogero-type systems to the
black hole physics. As was argued in  \cite{GibT},
${\cal N}=4$ supersymmetric extension of the conformal Calogero model can provide
a microscopic description of the extreme Reissner-Nordstr\"{o}m black hole in the near-horizon limit.
At the same time, the corresponding physical states are identified with the eigenstates of the Calogero-Moser Hamiltonian.
The deformed ${\cal N}=4$ supersymmetric generalization of the Calogero-Moser system found here can shed more light
on these issues. One can expect, e.g., that this new multiparticle ${\rm SU}(2|1)$ model exhibits a trigonometric
realization of the $d=1$ superconformal group $D(2,1;\alpha)$ along the lines of refs. \cite{Pop,HTop,IST}.

Finally, it is worth pointing out that we have obtained ${\cal N}=4$ supersymmetric extension of the $A_{n-1}$ Calogero-Moser system
by dealing with the matrix model with the ${\rm U}(n)$ gauging.
Superextensions of the Calogero-Moser models corresponding to other root systems could presumably
be obtained by choosing other gauge groups and/or representations for the matrix and WZ superfields.

\section*{Acknowledgements}

\noindent
We are indebted to Stepan Sidorov for interest in this work and useful discussions.
This research was supported by the Russian Science Foundation Grant No.\,16-12-10306.

\begin{appendices}
\section{Master ${\rm SU}(2|1)$ harmonic formalism}
\subsection{Extended harmonic setting}\label{App. A.1}
The formalism below is very similar to the bi-harmonic approach developed in \cite{GIO} for the harmonic space description
of quaternion-K\"ahler manifolds. The difference is that in \cite{GIO} all three extra co-ordinates $z^0, z^{\pm\pm}$ were introduced, while
in our case it will be enough to deal with two such coordinates $z^{\pm\pm}$.

Let us consider an extended ${\rm SU}(2|1)$ harmonic superspace in the $w$-parametrization of harmonic variables
\bea
(t_A, \theta^{\pm}, \bar\theta^{\pm}, w^\pm_i, z^{++}, z^{--}) = (\hat\zeta_H, z^{--})\,,\lb{A1}
\eea
where $z^{--}$ is an additional coordinate with the following ${\rm SU}(2|1)$ transformation properties
\bea
\delta z^{--} = \lambda^{--} - 2\lambda^{+-} z^{--}\,, \quad \lambda^{--} = m(\eta^-\bar\theta^- + \bar\eta^-\theta^-)\,, \;
\lambda^{+-} = m(\eta^-\bar\theta^+ + \bar\eta^-\theta^+) \,.\lb{Tranz--}
\eea
All other coordinates are transformed as in Sect.\,2. We assume that only generators ${I}^0$ and ${F}$ form the stability subgroup and
hence correspond to the homogeneous transformations of coordinates. Respectively, the general superfield given on \p{A1}, $\Phi( t, \theta, w, z,)\,,$ is assumed
to transform as (we consider passive transformations)
\bea
\delta \Phi = -\lambda^{+-} \tilde{I}^0 \Phi + 2\omega^{+-}\tilde{F} \Phi\,, \quad \omega^{+-} = m(\eta^-\bar\theta^+ - \bar\eta^-\theta^+)\,, \lb{DeltaPhi}
\eea
where $\tilde{I}^0$ and $\tilde{F}$ are just the ``matrix parts'' of the ${\rm U}(1)$ generators ${I}^0$ and ${F}$ counting two independent external ${\rm U}(1)$ charges
of $\Phi$. For sake of brevity we do not indicate these two charges explicitly. In general, $\Phi$ possesses also the standard harmonic ${\rm U}(1)$ charge $q$,
\bea
{\cal D}^0 \Phi &=& q\, \Phi\,, \nn
{\cal D}^0 &=& D^0_w + 2 z^{++}\frac{\partial}{\partial z^{++}} - 2 z^{--}\frac{\partial}{\partial z^{--}}\,, \nn
D^0_w &=& \partial^0_w + \theta^+\frac{\partial}{\partial \theta^+} +  \bar\theta^+\frac{\partial}{\partial \bar\theta^+}
-\theta^-\frac{\partial}{\partial \theta^-} - \bar\theta^-\frac{\partial}{\partial \bar\theta^-}\,.\lb{calD}
\eea

The covariant derivatives are defined by the following formulas
\bea
{\cal D}^{++}_{z} &=& {D}^{++}_w - (z^{++})^2\frac{\partial}{\partial z^{++}} + z^{++}({\cal D}^0 +{\tilde I}^0)  + [1 + m(\theta^+\bar\theta^-
- \theta^-\bar\theta^+)]\frac{\partial}{\partial z^{--}}\,, \nn
{D}^{++}_w &=& \partial^{++}_w + 2i \theta^+\bar\theta^+ \partial_t + \theta^+\frac{\partial}{\partial \theta^-} + \bar\theta^+\frac{\partial}{\partial \bar\theta^-} \nn
&&\, + \, m \theta^+ \bar\theta^+ \left(\theta^-\frac{\partial}{\partial \theta^-} - \bar\theta^-\frac{\partial}{\partial \bar\theta^-} \right)
- 2m \theta^+\bar\theta^+ \tilde{F}, \lb{D++1} \\
{\cal D}^{--}_{z} &=& {D}^{--}_w +[1 + m(\theta^+\bar\theta^-
- \theta^-\bar\theta^+)]\frac{\partial}{\partial z^{++}} - (z^{--})^2 \frac{\partial}{\partial z^{--}} + z^{--}\tilde{I}^0\,, \nn
{D}^{--}_w &=& [1 + m(\theta^+\bar\theta^- - \theta^-\bar\theta^+)]\partial_w^{--}+ 2i \theta^-\bar\theta^- \partial_t +\theta^-\frac{\partial}{\partial \theta^+}
+ \bar\theta^-\frac{\partial}{\partial \bar\theta^+}- 2m \theta^-\bar\theta^- \tilde{F}\,,  \lb{D--1} \\
{\cal D}^+ &=& \frac{\partial}{\partial \theta^-} - m \bar\theta^-\frac{\partial}{\partial z^{--}}\,, \quad \bar{\cal D}^+ =
-\frac{\partial}{\partial \bar\theta^-} + m \theta^-\frac{\partial}{\partial z^{--}}\,. \lb{D+1}
\eea
One should add to this set two more independent covariant derivatives
\be
\frac{\partial}{\partial z^{--}}\,, \quad \frac{\partial}{\partial z^{++}}\,, \quad \delta \frac{\partial}{\partial z^{--}} =
2 \lambda^{+-}\frac{\partial}{\partial z^{--}}\,, \;
\delta \frac{\partial}{\partial z^{++}} = 0\,.\lb{nablas}
\ee

It is also easy to define the covariant spinor derivatives ${\cal D}^-$ and $\bar{\cal D}^-$,
\bea
{\cal D}^-_z := [{\cal D}^{--},{\cal D}^+]\,, \quad \bar{\cal D}^-_z := [{\cal D}^{--}, \bar{\cal D}^+]\,. \lb{D-1}
\eea
For brevity, we will not present here their explicit form.

Now it is direct to be convinced that the quantities
\bea
{\cal D}^{\pm\pm}_z\Phi\,, \quad \frac{\partial}{\partial z^{\pm\pm}}\Phi\,, \quad {\cal D}^{+}\Phi\,, \quad \bar{\cal D}^{+}\Phi
\eea
(as well as ${\cal D}^{-}_z\Phi\,, \,\bar{\cal D}^{-}_z\Phi$) transform according to the generic superfield transformation law \p{DeltaPhi},
with taking into account that the covariant derivatives \p{D++1} - \p{nablas} themselves possess
non-trivial $\tilde{I}^0$ and $\tilde{F}$ charges\footnote{And of course the standard harmonic ${\rm U}(1)$ charges in accord with the numbers of $+$ and $-$ indices.}
\bea
&&\tilde{I}^0 \left({\cal D}^{++}_z, \,{\cal D}^{--}_z,\, {\cal D}^{+}, \, \bar{\cal D}^{+}, \,\frac{\partial}{\partial z^{++}},\, \frac{\partial}{\partial z^{--}}\right)
= \left(0, \,2\, {\cal D}^{--}_z,\, -\,{\cal D}^{+}, \, -\,\bar{\cal D}^{+}, \, 0,\, -2\,\frac{\partial}{\partial z^{--}}\right), \nn
&&\tilde{F}\left({\cal D}^{++}_z, \,{\cal D}^{--}_z,\, {\cal D}^{+}, \, \bar{\cal D}^{+}, \,\frac{\partial}{\partial z^{++}},\, \frac{\partial}{\partial z^{--}}\right)
= \left(0, \,0, \, -\frac12\,{\cal D}^{+}, \, \frac12\,\bar{\cal D}^{+}, \, 0, \, 0\right). \lb{Assign}
\eea

Note the useful (anti)commutation relations
\bea
&&\{{\cal D}^+, \bar{\cal D}^+\} = 2m\, \frac{\partial}{\partial z^{--}}\,, \quad [{\cal D}^{++}_z, {\cal D}^+] = [{\cal D}^{++}_z, \bar{\cal D}^+] = 0\,, \quad
[{\cal D}^{++}_z, {\cal D}^{--}_z] = 0\,, \lb{Comm1} \\
&& [\frac{\partial}{\partial z^{++}}, \,{\cal D}^{++}_z] = {\cal D}^0 + \tilde{I}^0\,,\quad [\frac{\partial}{\partial z^{++}}, \,{\cal D}^{--}_z] = 0\,, \nn
&&[\frac{\partial}{\partial z^{--}}, \, {\cal D}^{++}_z] = 0\,, \quad [\frac{\partial}{\partial z^{--}}, \,{\cal D}^{--}_z] = \tilde{I}^0\,. \lb{Comm2}
\eea
Defining
\bea
{\cal D}^{\pm\pm} = {\cal D}^{\pm\pm}_z -\frac{\partial}{\partial z^{\mp\mp}}\,,
\eea
we also find
\bea
[{\cal D}^{++}, {\cal D}^{--}] = {\cal D}^0\,.
\eea
While checking \p{Comm1}, \p{Comm2}, one should take into account the matrix ${\rm U}(1)$ charges assignment \p{Assign}. Also note that the ${\rm SU}(2|1)$
transformations of objects ${\cal D}^{\pm\pm}\Phi$, as distinct from ${\cal D}^{\pm\pm}_z\Phi$, reveal some deviations from the generic superfield law
\p{DeltaPhi}. For instance, ${\cal D}^{++}\Phi$, with ${\tilde I}^0 \Phi = p \Phi, \;{\tilde F}\Phi = l\Phi$, transforms as
\bea
\delta{\cal D}^{++}\Phi = -\lambda^{+-}p\,{\cal D}^{++}\Phi + 2\,\omega^{+-}l\,{\cal D}^{++}\Phi  -2 \lambda^{+-}\frac{\partial}{\partial z^{--}}\Phi\,.
\eea

\subsection{Eliminating $z$ dependence}\label{App. A.2}
We wish to deal with the superfields containing no dependence on the extra coordinates $z^{\pm\pm}$. As the first step,
we impose the manifestly covariant conditions
\bea
\mbox{a)}\;\;({\cal D}^0 + \tilde{I}^0)\Phi = 0\,, \quad \mbox{b)}\;\;\frac{\partial}{\partial z^{++}}\Phi = 0\,,\lb{Constrz++}
\eea
which eliminate the dependence on $z^{++}$ from both the superfield $\Phi$ and covariant derivatives\footnote{In some cases there is no need to impose (\ref{Constrz++}a),
still dealing with the $z^{++}$-independent superfields (see footnote on p.9).}. Now
\bea
&& \Phi \rightarrow \Phi(t, \theta, w, z^{--}) =:\Phi_{(z)} \,, \quad {\cal D}^{++}_z \rightarrow {D}^{++}_w +[1 + m(\theta^+\bar\theta^-
- \theta^-\bar\theta^+)]\frac{\partial}{\partial z^{--}}\,, \nn
&&{\cal D}^{--}_z \rightarrow {D}^{--}_w- (z^{--})^2 \frac{\partial}{\partial z^{--}} + z^{--}\tilde{I}^0\,, \quad {\cal D}^0 \rightarrow D^0_w
-2z^{--}\frac{\partial}{\partial z^{--}}\,. \lb{Subst}
\eea

Eliminating $z^{--}$ dependence is more subtle and admits three different possibilities. Before explaining this, let us pass to another form of the
transformation law \p{DeltaPhi} for $\Phi_{(z)}$, such that it is chosen to be active with respect to $\delta z^{--} = \lambda^{--} - 2\lambda^{+-}z^{--}$
\bea
\hat\delta \Phi_{(z)} = \lambda^{+-}D^0_w\Phi_{(z)} + 2\omega^{+-}\tilde{F}\Phi_{(z)} - \lambda^{--} \frac{\partial}{\partial z^{--}}\Phi_{(z)}\,,\lb{deltaAct}
\eea
where we made use of  \p{Subst} and the constraint (\ref{Constrz++}a).\\

Now we are prepared to discuss three options for eliminating $z^{--}$ dependence.\\

\noindent{\bf I}. The simplest possibility is to put
\bea
&& \frac{\partial}{\partial z^{--}}\Phi_{(z)} = 0\,, \quad \Phi_{(z)} \quad \Rightarrow \quad \phi(t, \theta, w)\,, \quad \hat\delta \phi = \lambda^{+-}D^0_w\phi
+2\, \omega^{+-}\tilde{F}\phi\,, \nn
&&{\cal D}^{++}_z \quad \Rightarrow \quad {D}^{++}_w\,, \quad {\cal D}^{--}_z \quad  \Rightarrow \quad {D}^{--}_w- z^{--}D^0_w\,, \quad {\cal D}^0 \rightarrow D^0_w\,.
\eea
In this case ${\cal D}^+ = \frac{\partial}{\partial \theta^-}\,, \quad \bar{\cal D}^+ = -\frac{\partial}{\partial \bar\theta^-}$ and one can impose
the ${\rm SU}(2|1)$ covariant Grassmann analyticity conditions $\frac{\partial}{\partial \theta^-}\phi = \frac{\partial}{\partial \bar\theta^-}\phi = 0$
without any need for the constraint ${D}^{++}_w\phi = 0$, as opposed to the harmonic formalism of \cite{IS15},
in which Grassmann analyticity conditions imply
the vanishing of the $++$ harmonic derivative of the analytic superfield. We also note that the action of the second covariant harmonic derivative
${\cal D}^{--}_z$ on $\phi$ produces a
superfield with a linear dependence on $z^{--}$,  ${\cal D}^{--}_z\phi = {D}^{--}_w\phi - z^{--}D^0_w\phi$,  unless $D^0_w\phi = 0$. Correspondingly, ${D}^{--}_w\phi$
transforms through the superfield $\phi$ itself.  One can show that the same subtleties take place for the spinor derivatives ${\cal D}^-\phi$ and $\bar{\cal D}^-\phi$.\\

\noindent{\bf II}. The harmonic formalism of \cite{IS15} is recovered,  when the $z^{--}$ dependence of $\Phi_{(z)}$ is  fixed in a more sophisticated way, by imposing
the constraint
\bea
{\cal D}^{++}_z\Phi_{(z)} = 0 \rightarrow  \frac{\partial}{\partial z^{--}}\Phi_{(z)} = -[1 + m(\theta^+\bar\theta^- - \theta^-\bar\theta^+)]^{-1}{D}^{++}_w\Phi_{(z)}\,.
\lb{Bcond}
\eea
This condition expresses all the coefficients in the $z^{--}$ power series expansion of $\Phi_{(z)} = \phi(t,\theta, w) + z^{--}\phi^{++}(t,\theta, w) + \ldots$
in terms of powers of $\tilde{D}^{++}_w := [1 + m(\theta^+\bar\theta^- - \theta^-\bar\theta^+)]^{-1}{D}^{++}_w$ acting on the lowest coefficient, i.e. on $\phi$.
The transformation law \p{deltaAct} is reduced to
\bea
\hat\delta \phi = \lambda^{+-}D^0_w\phi + 2\,\omega^{+-}\tilde{F}\phi + \lambda^{--} \tilde{D}^{++}_w\phi\,, \lb{Btrnasflaw}
\eea
that is precisely the generic superfield ${\rm SU}(2|1)$ transformation law postulated in \cite{IS15}. The harmonic derivatives  $\tilde{D}^{++}_w$ and ${D}^{--}_w$ coincide
with those defined in \cite{IS15}, $[\tilde{D}^{++}_w, \, {D}^{--}_w]  = D^0_w$. The objects $ \tilde{D}^{++}_w\phi$, ${D}^{--}_w\phi$ and
${\cal D}^+\phi = (\frac{\partial}{\partial \theta^-} + m\bar\theta^-\tilde{D}^{++}_w)\phi\,$, $\bar{\cal D}^+\phi = (-\frac{\partial}{\partial \bar\theta^-} -
m\theta^-\tilde{D}^{++}_w)\phi$ are transformed according to \p{Btrnasflaw} \footnote{The same is true for the $z$-independent parts of the covariant
spinor derivatives ${\cal D}^-, \bar{\cal D}^-$ in which the substitution \p{Bcond} has been made.}. The harmonic Grassmann analyticity for $\phi$ implies
the constraint $\tilde{D}^{++}_w\phi = 0$. \\

\noindent{\bf III}. Yet one more way to fix the $z^{--}$ dependence of $\Phi_{(z)}$ is to impose the condition like the well-known Scherk-Schwarz reduction condition
\bea
&&\Phi_{(z)} = {e}^{z^{--}\tilde{I}^{++}}\,\phi'(t,\theta, w)\,, \quad \frac{\partial}{\partial z^{--}}\Phi_{(z)} =
{e}^{z^{--}\tilde{I}^{++}}\,(\tilde{I}^{++}\phi')\,, \quad [\tilde{I}^0, \tilde{I}^{++}] = 2\tilde {I}^{++}\,, \lb{Cconstr} \\
&&\hat\delta\phi' = \lambda^{+-}D^0_w\phi' + 2\,\omega^{+-}\tilde{F}\phi' - \lambda^{--} \tilde{I}^{++}\phi'\,. \lb{Ctransflaw}
\eea
The corresponding version of the ${\rm SU}(2|1)$ harmonic formalism is just the one constructed and discussed in Sect.\,2. In particular, ${\cal D}^{++}_z = {\cal D}^{++}_w
+ \tilde{I}^{++}$,
where ${\cal D}^{++}_w$ is now just \p{cov-D++} written in the $(w^\pm_i, z^{++})$ basis and restricted to the superfields satisfying the conditions \p{Constrz++}
\footnote{Actually, the condition \p{Cconstr} can be imposed {\it before} \p{Constrz++}, so that ${\cal D}^{++}_w$ will precisely coincide with \p{cov-D++} in
the $(w^\pm_i, z^{++})$ basis.}.
The covariant derivative ${\cal D}^{--}_z$ defined in \p{D++} coincides,  on the same subclass of ${\rm SU}(2|1)$ superfields, with ${D}^{--}_w$. Actually, the option {\bf III}
is very similar to the option {\bf I}. Like in the latter case, the Grassmann analyticity requires $\tilde{I}^{++}\phi = 0$, but not $\tilde{D}^{++}_w\phi=0$ as in \cite{IS15}.

\end{appendices}

\end{document}